\documentclass[11pt,a4paper]{article}
\pdfoutput=1
\usepackage{latexsym}
\usepackage{graphics,graphicx}
\usepackage{amsmath,amsfonts,amssymb,indentfirst,mathrsfs,bm,fancyhdr}
\usepackage[utf8]{inputenc}
\usepackage[T1]{fontenc}
\usepackage{esint}
\usepackage{epsfig}
\usepackage{cite}
\usepackage{color}
\usepackage{float}

\newcommand{\be}{\begin{equation}}
\newcommand{\ee}{\end{equation}}

\def\bea{\begin{eqnarray}}
\def\eea{\end{eqnarray}}

\def\beqx{\begin{displaymath}}
\def\eeqx{\end{displaymath}}

\newcommand{\bmat}{\left(\begin{array}}
\newcommand{\emat}{\end{array}\right)}

\def\i{\iota}

\def\bo{{\raise-.3ex\hbox{\large$\Box$}}}               % D'Alembertian
\def\face{{\raise.2ex\hbox{$\displaystyle \bigodot$}\mskip-2.2mu \llap {$\ddot
        \smile$}}}                                   % happy face
\def\>{\rangle}                                      %right angle
\def\<{\langle}                                      %left angle

\def\leftrightarrowfill{$\mathsurround=0pt \mathord\leftarrow \mkern-6mu
        \cleaders\hbox{$\mkern-2mu \mathord- \mkern-2mu$}\hfill
        \mkern-6mu \mathord\rightarrow$}        % <--> double differential
\def\dvec#1{\vbox{\ialign{##\crcr
        \leftrightarrowfill\crcr\noalign{\kern-1pt\nointerlineskip}
        $\hfil\displaystyle{#1}\hfil$\crcr}}}           % <--> accent
\def\-{\hphantom{-}}

\title{\vskip -0.5in
\begin{minipage}{6.5in}
\begin{flushright}
 {\small IFT-UAM/CSIC-14-031}
\end{flushright}
\end{minipage}\\
\vskip 1.0in
Anomalous Transport in Holographic Chiral Superfluids via\\ Kubo Formulae
}

\author{Amadeo Jimenez-Alba\footnote{amadeo.j@gmail.com}\,\, and\, Luis Melgar\footnote{luis.melgar@csic.es}\\ \\
{\it Instituto de F\'{\i}sica Te\'orica UAM/CSIC} \\ {\it Universidad Aut\'onoma de Madrid}\\ {\it Madrid 28049, Spain}}

\begin{document}

\maketitle

\begin{abstract}
We study anomalous conductivities in Chiral Superfluids in the framework of two different holographic models, by means of Kubo formulae. In addition, we point out the existence of an anomalous transport phenomenon that consists in the presence of a charge density when the superfluid velocity is aligned with a magnetic field. It has been pointed out recently that certain chiral conductivities in holographic superfluids exhibit universal behavior at zero temperature. We show that anomalous conductivities always stabilize at low temperatures in our setup. Even though the particular value they acquire is model-dependent, it seems to be robust and determined solely by the interplay between the broken symmetries and the anomalies. 
\end{abstract}

\newpage

%\tableofcontents
%\addtocontents{toc}{\protect\setcounter{tocdepth}{3}}
\renewcommand{\theequation}{\arabic{section}.\arabic{equation}}

\section{Introduction}
During the last few years, there has been a increasing interest in the transport phenomena driven by anomalies of the microscopic Quantum Field Theory (QFT)\cite{Bertlmann:1996xk}. Several studies have been carried out from the point of view of both Hydrodynamics \cite{Neiman:2010zi,Sadofyev:2010pr,Son:2009tf,Haehl:2013hoa,Banerjee:2013fqa,Avdoshkin:2014gpa} and Kubo Formulae \cite{Kharzeev:2009pj, Landsteiner:2011cp, Landsteiner:2012kd}. Due to anomalies, ordinary fluids respond to the presence of an external magnetic field $\vec B$ or vorticity $\vec \omega$ generating a current in the direction of the external sources. This has been called the Chiral Magnetic Effect (CME) and Chiral Vortical Effect (CVE), respectively. The associated conductivities are the Chiral Magnetic and Chiral Vortical Conductivities (CMC and CVC respectively). These coefficients have been computed and understood from many different approaches and perspectives, both in QFT \cite{Kharzeev:2007tn,Fukushima:2008xe,Loganayagam:2011mu,Jensen:2012jy, Banerjee:2012cr,Loganayagam:2012pz} and holography \cite{Erdmenger:2008rm, Banerjee:2008th,Newman:2005hd,Yee:2009vw, Amado:2011zx,Landsteiner:2011iq,Jensen:2012kj,Kalaydzhyan:2011vx} (for a recent review, see \cite{Pena-Benitez:2013mma}). \\
It could be stated that the study of the interplay between anomalous transport and superfluids started a decade ago; the first approaches to chiral transport (concretely, the Chiral Separation Effect) where analyzed for high-density QCD, assuming for instance that baryon symmetry is spontaneously broken, see \cite{Metlitski:2005pr,Son:2004tq}. However, a systematic study of Chiral Superfluids has only been undertaken recently, using different techniques to obtain the hydrodynamic expansion, with particular emphasis on the anomalous response \cite{Lublinsky:2009wr,Lin:2011aa, Lin:2011mr, Neiman:2011mj,Bhattacharyya:2012xi}. \\
The results indicate that the effect of the background condensate is two-fold. On the one hand, unlike the case of ordinary fluids, anomalous conductivities are not fully determined by anomaly coefficients anymore. On the other, in addition to the Chiral Vortical and Chiral Magnetic effects, there exist new types of transport phenomena driven by the anomalies. However, until now, we lack clear predictions for the anomalous response parameters in superfluids.\\
Remarkably, it has been recently pointed out that, for a certain class of holographic models of chiral superfluids \cite{Bhattacharya:2011tra} the zero-temperature behaviour of the CMC and CVC is universal and given by \cite{Amado:2014mla} 
\begin{align}
\label{Amadoetal1} \sigma^{\text{brok.}}_{55} (T\rightarrow 0) &= \frac{  \sigma^{\text{unbrok.}}_{55} }{3}\,,\\
\label{Amadoetal2} \sigma^{\text{brok.}}_{CVC} (T\rightarrow 0) &= 0\,,
\end{align}
where "brok." and "unbrok." refer to broken and unbroken phases, respectively.\\

Here we address possible corrections of the anomalous transport coefficients due to the presence of condensates, performing an explicit computation of them. We will focus on the strongly coupled regime and, to simplify the approach, we will stick to s-wave condensates, which correspond to broken phases in which the order parameter is a scalar. To that end, we use holographic methods.  \\
Contrary to the usual approaches to transport in Chiral Superfluids, here we will rely on linear response theory to analyze the possible corrections. Kubo formulae provide us with the response driven by a small external perturbation. These are powerful because they account automatically for all the corrections to the coefficients and sometimes prove the existence of new transport phenomena which is difficult to analyze by means of hydrodynamic expansions. 
Hence, we assume that it is possible to define the anomalous conductivities in terms of correlators in the broken phase, which is to say, that there exists a current due to an external magnetic field in both the unbroken and broken phases\footnote{For a detailed analysis of some of the Kubo formulae applied to Chiral Superfluids, see \cite{Chapman:2013qpa}}
\begin{align}
\label{Chiral1}\mathcal{J}^i &= \sigma_{\{CME, CSE, 55\}} B^i\,,\\
\label{Chiral2}\mathcal{J}^i &= \sigma_{CEE} \epsilon^{ij}E_j\,.
\end{align} 
Where $\mathcal{J}, B, E$ correspond to a generic $U(1)$ covariant current, magnetic and electric field, respectively, whereas $\sigma$ denotes generic conductivities\footnote{CSE stands for Chiral Separation Effect.}. Equation (\ref{Chiral2}) represents the Chiral Electric Effect (CEE), an anomalous transport phenomenon which is present only for Chiral Superfluids at finite superfluid velocity \cite{Neiman:2011mj}. We will propose a Kubo formula for the Chiral Electric Conductivity (CEC) and compute its value in our models.  \\
In addition, we emphasize the existence of a type of transport phenomena in Chiral Superfluids that to our knowledge has been overlooked so far. We will call it Chiral Charge Generation Effect (CCGE). It establishes the presence of a charge density whenever the supervelocity is aligned with an external magnetic field
\begin{align}
\label{Chiral 3}\rho = \hat{\sigma}\  \vec \xi \cdot  \vec B
\end{align}
here $\vec \xi$ is the superfluid velocity and $\hat{\sigma}$ the corresponding conductivity (CCGC). We will provide a Kubo formula for it in Section \ref{sec:Kubo} and compute its value, showing that it is generically different from zero. The response prescribed by (\ref{Chiral 3}) is not formally new, even though we believe its physical importance has not been stressed before. It has appeared in the literature and for instance it can be mapped to the term $S_1$ of equation (2.31) of \cite{Bhattacharyya:2012xi}  \footnote{We thank Carlos Hoyos for pointing this out. See also \cite{Kalaydzhyan:2012ut}}. Such a term establishes the presence of a charge density whenever a transverse London-type-current $S_1 = \epsilon^{ijk}\zeta_i\partial_j \zeta_k$ is acting on the system. Since $\zeta_k = -\partial_k \phi + A_k$ (see \cite{Bhattacharyya:2012xi}) we propose that there is an effective response of the form (\ref{Chiral 3}) arising from $S_1 = \epsilon^{ijk}\zeta^0_i\partial_j A_k +...$. We believe that such a transport phenomenon leads to interesting phenomenological implications. \\ 

Notice that, for the above formulae to make sense, it is important in general that the background we are considering is stable in the presence of a (perturbatively small) magnetic field, i.e. that there exists a perturbative expansion in the amplitude of a external magnetic field. Given such a perturbative expansion, at zeroth order the holographic superfluid corresponds to the background considered here. This is consistent with the usage of Kubo Formulae to compute the transport coefficients. However, for finite external magnetic fields, the holographic superfluid gets affected and, in particular, it generates London-type currents\cite{Hartnoll:2008kx}. Therefore, one could argue against the validity of our results beyond perturbatively small external sources. In order to avoid that potential issue, in Section \ref{sec:axialvec} we study a $U(1)\times U(1)$ model, in which only one of the $U(1)$'s undergoes a phase transition and thus we can study how the (unscreened) magnetic field associated to the unbroken symmetry enters the chiral transport properties.\\
In what follows we will work with global symmetries in the QFT, for they are very naturally accommodated within holography. This means that we can restrict ourselves to configurations which do not excite the anomaly. This is a pertinent remark, since having a dynamical photon would imply the existence of general loop corrections to the anomalous transport coefficients \cite{Jensen:2013vta} which are important even in the hydrodynamic approximation. Despite the fact that there is no photon here, in the broken phase the Goldstone boson could in general give important corrections at strong coupling. However, we expect our calculation not to capture all these contributions, for they are subleading in the classical gravity approximation.\\
A source of the corrections that we should be able to capture within holography is the one associated to the background scalar field. For instance, in \cite{Newman:2005as} the Chiral Separation Conductivity (CSC) was indeed found to present corrections in the case of a linear sigma model (the background scalar field gives an effective mass to the fermions through the Yukawa coupling and contributes to the CSC).\\

Entropic arguments were used in \cite{Neiman:2011mj} to extract the Hydrodynamics of Chiral Superfluids in the presence of external unbroken gauge fields. The Chiral Electric Effect was predicted and some possible generic corrections to the CMC and CVC were found. Moreover, in \cite{Amado:2014mla} it was argued that such corrections do not vanish but become universal (model independent) at low temperatures and the CMC and the CVC were computed at $T=0$, indeed finding a universal result. Our models are restricted to the probe limit and hence we will not be able to reach $T\rightarrow 0$; furthermore, we cannot induce metric perturbations and hence the CVC cannot be calculated. However, we observe that the chiral conductivities stabilize fast enough to be able to observe their $T=0$ behaviour even at temperatures close enough to $T_c$, where our computations are reliable.\\
In what follows we consider two models, one in which a $U(1)$ anomalous symmetry undergoes a phase transition and one in which we have two $U(1)$ symmetries and only one of them develops a condensate. In the absence of supervelocity the former case reduces to a truncation of the model of \cite{Amado:2014mla} and indeed we observe that $\sigma_{55}$ approaches the value prescribed by equation (\ref{Amadoetal1}). In the latter model (not considered so far in the literature)  we can define three non-vanishing anomalous conductivities at zero supervelocity \cite{Gynther:2010ed}; our results suggest that all of them approach universal values at low temperatures. Remarkably enough, the universal ratio is always different from  $1/3$ and, in particular, the CMC vanishes as we increase the chemical potential. 

\subsection{Remarks on the definition of the current}
At this point it is important to point out several remarks related to the definition of the currents. In principle, one could use the consistent currents to define the anomalous correlators. As pointed out in \cite{Rubakov:2010qi}, one has to be careful in this case, for the gauge fields at infinity are not directly related to the chemical potential of the theory. In \cite{Gynther:2010ed} an holographic calculation of the anomalous transport coefficients, taking the previous issue into account, was carried out; it was shown how one has to give up the condition that the background gauge field vanishes at the horizon in order to be able to distinguish the source from the chemical potential.\\
In the presence of a condensate, regularity imposes that the gauge field must be zero at the horizon. Hence, it is better to work from the start with the covariant definition of the current, as in \cite{Landsteiner:2012dm}. Notice that this amounts to neglecting the contribution to the current operator coming from the holographic Chern-Simons term. With this manipulation there is no trace of the sources in the correlators and one can perfectly work with a boundary condition such that the background gauge field vanishes at the horizon. The resulting correlators are the ones of \cite{Gynther:2010ed} with $\alpha=\beta=0$. Physically, we thus will be working with the covariant current\footnote{The covariant current is a gauge-invariant object and thus the source that couples to it is a good chemical potential. Therefore, by working from the begining with the covariant current we avoid the necessity of taking into account the difference between the source for the consistent current, $A_0$, and the actual (gauge-invariant) chemical potential, $\mu$ (see \cite{Gynther:2010ed} for a detailed discussion on this issue).}, and our computed retarded two-point functions contain therefore one covariant and one consistent current, namely
\begin{align}
\mathcal{G}_{\mathcal{R}}\sim \left<\mathcal{J}^{\text{cov}} \mathcal{J}^{\text{cons}}\right>\,.
\end{align}
Notice that this in particular implies that, no matter the model under consideration, none of our (covariant) currents is conserved in general. However, this is not a problem at all since our background gauge field configurations are such that the anomaly is not excited.

\subsection{Remarks on the Kubo Formulae}
\label{sec:Kubo}
Let us point out some remarks on the Kubo formulae we are going to use. We lack formal derivation of the one corresponding to the CEC. However, assuming a constitutive relation of the form (\ref{Chiral2}), we can derive a suitable Kubo formula for it. We point out that we do not intend to make contact with the hydrodynamic construction of \cite{Neiman:2011mj} (for example, our Kubo relations are associated to the laboratory frame, not the Landau frame). Instead, we will propose suitable Kubo formulae for the conductivities we aim to study, based on the fact that we know which the gauge-invariant sources are, as well as the type of response that we expect. Our Kubo formulae read
\begin{align} 
\label{C55} \sigma_{\{55,CSE,CME\}} =& \lim_{k\rightarrow 0} \frac{i}{2k} \left<J^y J^z\right>_{\mathcal{R}}(\omega=0, k)\,,\\
\label{CECaxial}\sigma_{CEC} =& \lim_{\omega\rightarrow 0} \frac{i}{2\omega} \left<J^y J^z\right>_{\mathcal{R}}(\omega, k=0)\,,\\
\label{CCGC} \sigma_{CCGE} =& \lim_{k\rightarrow 0} \frac{i}{2k_{\bot}} \left<J^0 J^y\right>_{\mathcal{R}}(\omega=0, k)\,.
\end{align}
Where $k_{\bot}$ means that the momentum points in a direction transverse to the supervelocity. All the conductivities in (\ref{C55}) are associated to similar correlators. The distinction between them comes from the nature of the currents inside the two point functions and it only makes sense in the presence of more than one $U(1)$. 
This will be made explicit in Section \ref{sec:axialvec}. We believe the above provide suitable expressions due to the following
\begin{itemize}
\item All the above conductivities vanish in the absence of anomaly.

\item For $\sigma_{\{55;CSE;CME\}}$ we rely on the fact that they are related to the response to an external magnetic field by definition. Moreover, as we will see, (\ref{C55}) is continuous through the phase transition, matching the value that $\sigma_{\{55,CSE,CME\}}$ shows in the unbroken phase. In addition to this, our formula coincides with the one of \cite{Chapman:2013qpa}.

\item In the case of $\sigma_{CEC}$, we take into account that it corresponds to the effect of an external electric field, as in \cite{Neiman:2011mj}. With this in mind, we choose a kinematic limit such that it can be drastically distinguished from the other anomalous transport coefficients. Moreover, we will observe that $\sigma_{CEC}\sim \xi$ at low temperatures.

\item The formula (\ref{CCGC}) can be derived from the discussion of  \cite{Chapman:2013qpa} (our notation is also taken from that reference). We start with the term $J^0 = -T_0 e^{\sigma} g_{1,\nu} S_1$ \footnote{$g_{1,\nu}$ is the derivative of the thermal coefficient $g_1$ with respect to $\nu \equiv \mu/T$ \cite{Bhattacharyya:2012xi}} and take the variation
\begin{align}
\frac{\delta S_1}{\delta A_l} =2i k_j \epsilon^{ijk} \zeta^{eq.}_i \frac{\delta \zeta^{eq.}_k}{\delta A_l} |_{\text{sources=0}}
\end{align} 
where the 2 comes from the fact that we have twice the same contribution $\epsilon^{ijk} \zeta^{eq.}_i\partial_j \frac{\delta \zeta^{eq.}_k}{\delta A_l}$. For transverse momentum, we use equation (3.29) of \cite{Chapman:2013qpa}, yielding
\begin{align}
\frac{\delta \zeta^{eq.}_i}{\delta A_l} =& \delta_i^l - \frac{k_i k^l}{k^2} - 2i T_0c_3 k_i \zeta^l_0\,,\\
\frac{\delta S_1}{\delta A_l} =&2i k_j \epsilon^{ijk} \zeta^{0}_i \left(\delta_k^l - \frac{k_k k^l}{k^2} - 2i T_0c_3 k_k \zeta^l_0\right) \,.
\end{align}
Now, $k_kk_j \epsilon^{kj} =0$ and hence, to first order in $k$ we have
\begin{align}
\left<J^0 J^l\right>= -2i  T g_{1,\nu} k_j \epsilon^{ijl} \zeta_i+ \mathcal{O}(k^2)
\end{align}
where all the equilibrium super/subscripts "0" have been omitted. From here, 
\begin{align}
\epsilon_{lmn} \mathcal{G}^{0l}_R = - 2i T g_{1,\nu}k_j \zeta_i \left(\delta_m^i\delta_n^j- \delta_m^j\delta_n^i\right)
\end{align}
where $\mathcal{G}^{0l}_R \equiv \left<J^0 J^l\right>$. The formula (\ref{Chiral 3}) can be recovered by assuming $m= z, n=x$. In our notation $\zeta_i \equiv \xi_i$ and we get\footnote{Notice in passing that the coefficient $g_1$, as defined in \cite{Bhattacharyya:2012xi}, is associated to a gauge-invariant term and hence cannot be fixed by anomaly matching. This makes the relation between chiral transport coefficients and underlying anomalies more subtle than in the case of ordinary fluids (see however Section \ref{subsec:Lowt}).}
\begin{align}
\label{CCGEKubo}\sigma_{CCGE} \equiv T \xi_z g_{1,\nu} =  \lim_{k_x \rightarrow 0}\frac{i}{2k_x} \mathcal{G}^{0y}_R (\omega=0)
\end{align}
\end{itemize}
To avoid any possible confusion let us point out that, taking advantage of the fact that we work with a fixed component of the supervelocity, throughout this work we will usually absorb the supervelocity factors into the conductivities, as prescribed by equations (\ref{CECaxial}) and (\ref{CCGC}). This can be seen explicitly in (\ref{CCGEKubo}). Of course, in general one has to take into account that the CEE and CCGE are linear in the supervelocity (a vector) and write expressions like (\ref{Chiral 3}) instead. \\
In Section \ref{sec:Brokaxial} we present a simple model in which we only have one $U(1)$ anomalous symmetry that gets broken spontaneously. We reproduce the outcomes of \cite{Amado:2014mla} and we also include finite supervelocity and analyze the results; in particular, we compute the CEC and the CCGC via Kubo formulae, showing that they do not vanish in general. Then we move to Section \ref{sec:axialvec}, where a more realistic model is considered: we work with a $U(1)\times U(1)$ symmetry, which can be interpreted as having both axial and vector currents (for a different interpretation, see the main text), with the condensate coupled to the vector sector. The richer set of chiral conductivities is analyzed (both at zero and finite supervelocity) with special emphasis on the $T\rightarrow 0$ behaviour suggested by data. Section \ref{sec:conc} includes interpretations, conclusions and future directions of the present work. 
\section{Broken Anomalous symmetry}
\label{sec:Brokaxial}
We want to analyze, from the holographic point of view, how anomalous conductivities are altered due to the presence of an s-wave condensate. To this end we consider a holographic superconductor plus a Chern-Simons term that induces a $U(1)^3$ anomaly in the dual field theory.\\
From the point of view of the dual field theory we have a spontaneously broken $U(1)$ anomalous global symmetry. The action of the bottom-up model reads
\begin{align} \label{action}
S= \int d^{5}x \sqrt{-g}\left(-\frac{1}{4} F_{MN}F^{MN} + \frac{\kappa}{3}\epsilon^{MABCD}A_MF_{AB}F_{CD} - \overline{D_{M}\Psi} D^M\Psi - m^2 \bar\Psi \Psi\right)
\end{align}
This is the model of \cite{Amado:2014mla} with $V_{\psi} =1$, $V = m^2 \bar\Psi \Psi $ and $\kappa = c/8$. In what follows we will be working with the covariant definition of the current, meaning that we are neglecting the Chern-Simons contribution to the definition of $J^{\mu}$.\\
We take the Schwarzschild AdS Black Brane in 5 dimensions as our background metric in the bulk
\begin{align}
ds^2= -f(r) dt^2 +\frac{dr^2}{f(r)} + \frac{r^2}{L^2}(dx^2+dy^2+dz^2)
\end{align}
being $f(r)= \frac{r^2}{L^2} - \frac{r^2_H}{r^2}$. From now on we will work in adimensional units, rescaling all the $L^2$ factors to one. Our ansatz for the background fields consists of a non-vanishing temporal and spatial component of the gauge field and the real component of the scalar field. All of them with just radial dependence
\begin{align}
A = \phi(r)dt + V (r)dx ; \hspace{2cm} \Psi(r) = \psi(r)
\end{align}
With this ansatz the background equations of motion reduce to
\begin{align}
\label{backg1} \phi'' + \frac{3}{r}\phi' - \frac{2 \psi^2}{f}\phi=0 \\ 
\label{backg2} \psi'' + \left(\frac{f'}{f}+ \frac{3}{r}\right) \psi' + \frac{\phi^2}{f^2}\psi - \frac{V^2}{r^2 f}\psi- \frac{m^2}{f} \psi =0\\
\label{backg3} V'' + \left(\frac{f'}{f}+ \frac{1}{r}\right) V'  - \frac{2 \psi^2}{f}V=0
\end{align}
The equations boil down to the ones which govern the usual s-wave holographic superconductor in the presence of supervelocity. This could have been anticipated by noticing that the ansatz does not excite the Chern-Simons contribution $\kappa \epsilon^{MABCD}F_{AB}F_{CD}$ to the gauge field equation. Hence, the anomaly is absent at the level of the background. However, it has important implications for the perturbations.  \\
\begin{figure}[h]
\centering
\includegraphics[width=230pt]{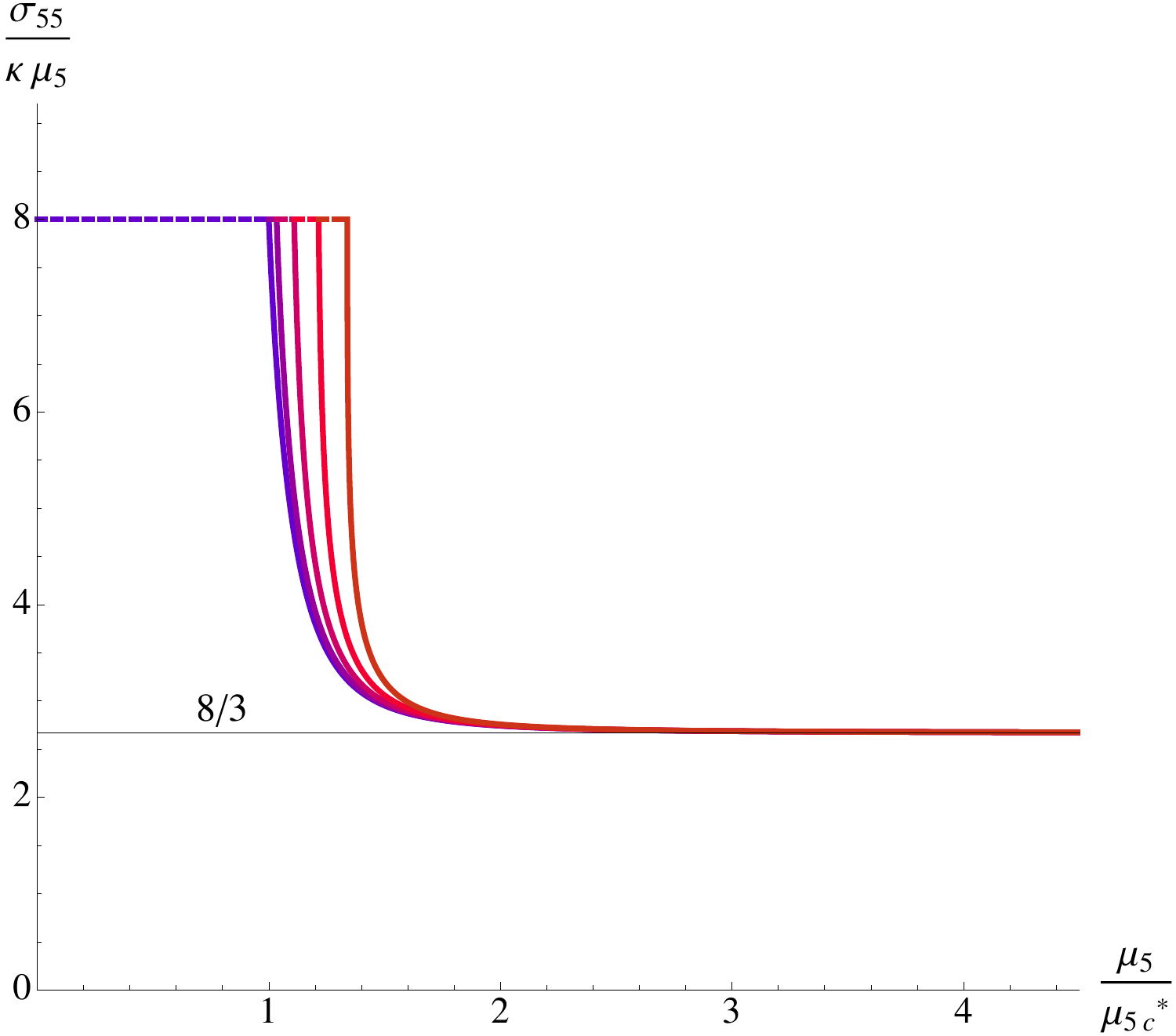}\hspace{2mm}
\includegraphics[width=230pt]{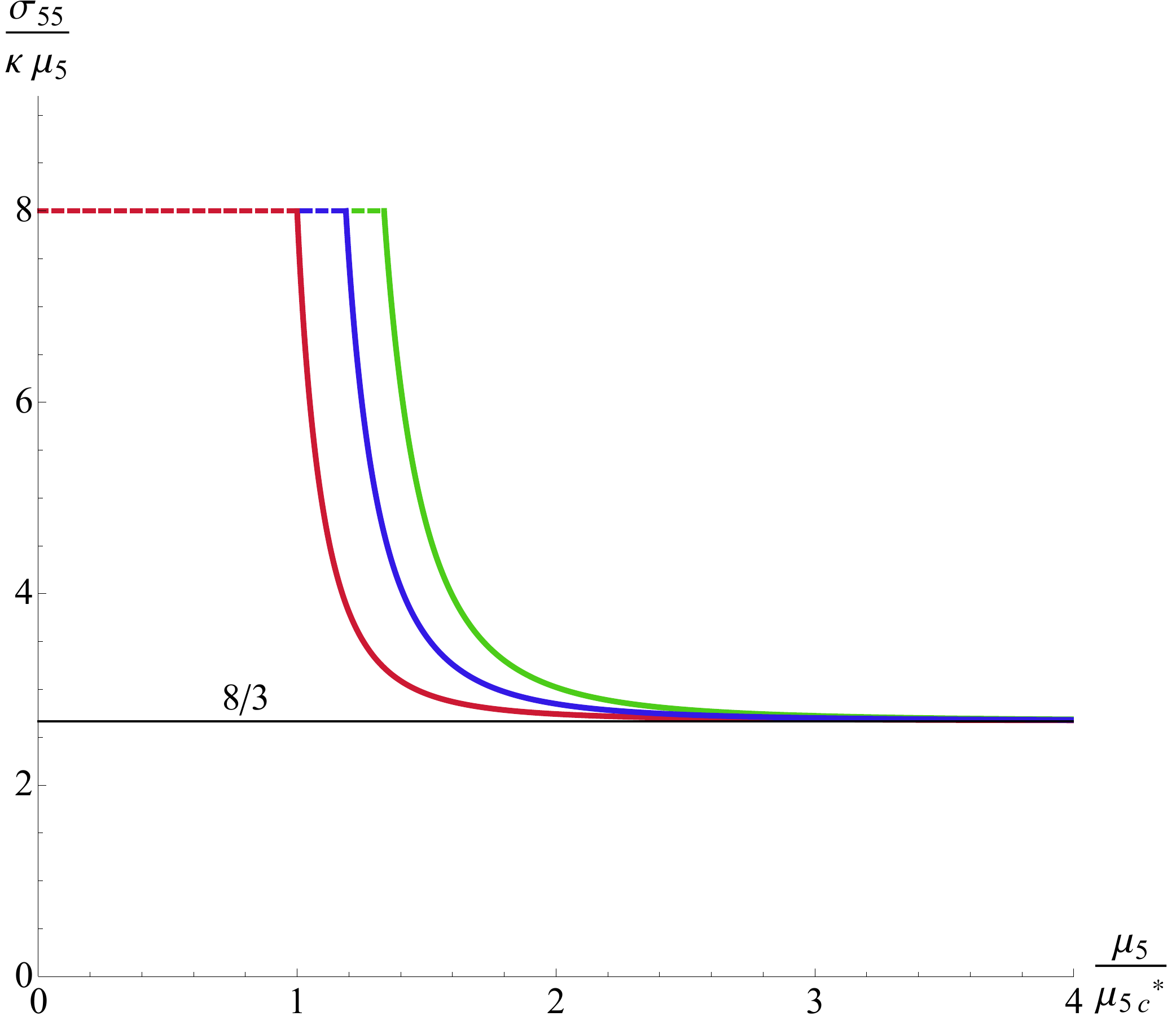}
\caption{\label{fig:Sigma55vsV} Axial conductivity divided by the chemical potential and the anomaly coefficient versus chemical potential. $\mu_{5c} ^*$ is the critical chemical potential at zero supervelocity . (Left) Each line corresponds to same mass value $m^2=-7/2$ and different superfluid velocity, from $\xi_x/T = 0.1$ (blue) to $\xi_x/T  =2.1$ (orange). The dashed horizontal line corresponds to the unbroken phase, where $\sigma_{55} \sim \mu_5$. In the broken phase this conductivity approaches $1/3$ of the unbroken phase value for large enough chemical potential. This is compatible with the results of \cite{Amado:2014mla}. (Right) Each line corresponds to a different mass (red $m^2=-7/2$, blue $m^2=-3$, green $m^2=-5/2$) of the scalar field in the bulk. As one can see the $1/3$ factor is unaltered by the dimension of the operator that condenses. The conductivity depends linearly with $\kappa$. }
\end{figure}
In our convention we choose to fix the temperature and interpret the adimensional quantities
\begin{equation}
\bar \mu = \frac{\mu }{T}; \hspace{2cm} \vec {\bar{\xi}} = \frac{\vec \xi}{T}
\end{equation}
as the chemical potential and the supervelocity of the system (along this work we make some abuse of language and refer to the $\bar \mu \rightarrow \infty$ regime as the $T\rightarrow 0$ limit), which are determined by the boundary conditions of the fields to be imposed at spatial infinity:
\begin{align}
\phi(r)_{r\rightarrow \infty}\sim \mu_5 \hspace{3cm} V(r)_{r\rightarrow \infty}\sim \xi_{\{x,z\}}
\end{align}
By $\xi_{\{x,z\}}$ we mean that the supervelocity will be taken to be pointing either in the $x$ or the $z$-direction. In addition, we choose the standard quantization, by imposing the boundary conditions to the leading term in the asymtotic expansion of the scalar field
\begin{align}
\nonumber \psi(r)_{r\rightarrow\infty}\sim \frac{\psi_1}{r^{\Delta_-}}+ \frac{\psi_2}{r^{\Delta_+}}+... \\\nonumber\\
\psi_1=0  \hspace{2.7cm}\psi_2 = \langle O \rangle
\end{align}
We solve equations (\ref{backg1})-(\ref{backg3}) with this boundary conditions numerically. \\

Before we proceed to discuss our results for the conductivities a comment is in order regarding the background we have constructed. The phase diagram of a holographic superconductor in presence of finite supervelocity was first studied in \cite{Herzog:2008he,Basu:2008st}. In a later study \cite{Amado:2013aea} it was shown that the system presents instabilities at finite momentum close to the phase transition for a large range of supervelocities. The stable background in that region is not known. Although this analysis was made in $AdS_{3+1}$ we expect it to apply in $AdS_{4+1}$ as well. We do not discard those issues to have some influence, even though, as we will see later on, all of our results seem to be perfectly consistent for every value of the chemical potential. In any case, let us emphasize that our forthcoming main observations have to do with the behaviour of the conductivities far from the transition point, where the above potential issues are not expected to play any role.
\subsection{The Chiral conductivities in the broken phase: Axial conductivity and CEC}
\begin{figure}[h] 
\centering
\includegraphics[width=230pt]{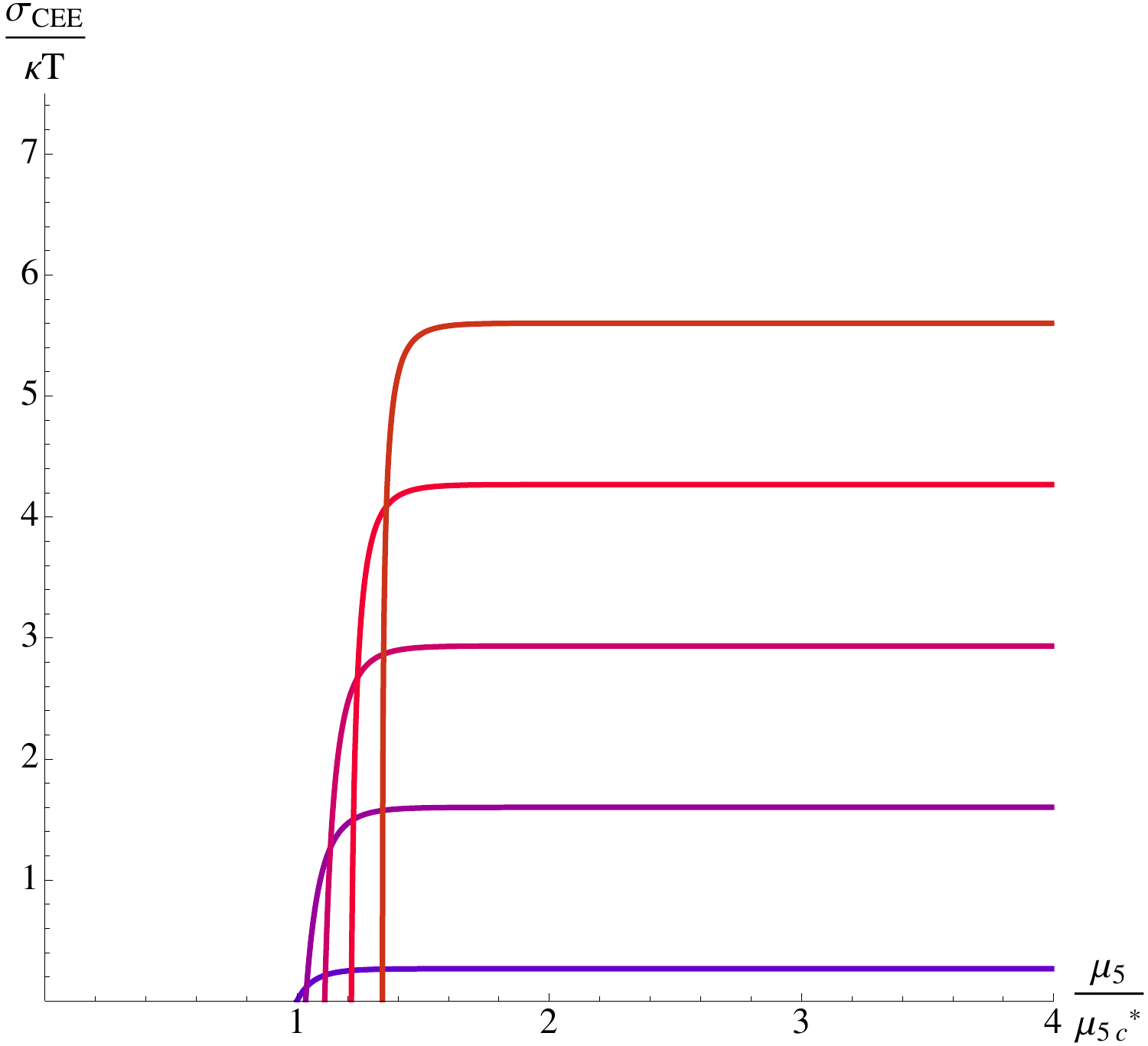}\hspace{2mm}
\includegraphics[width=230pt]{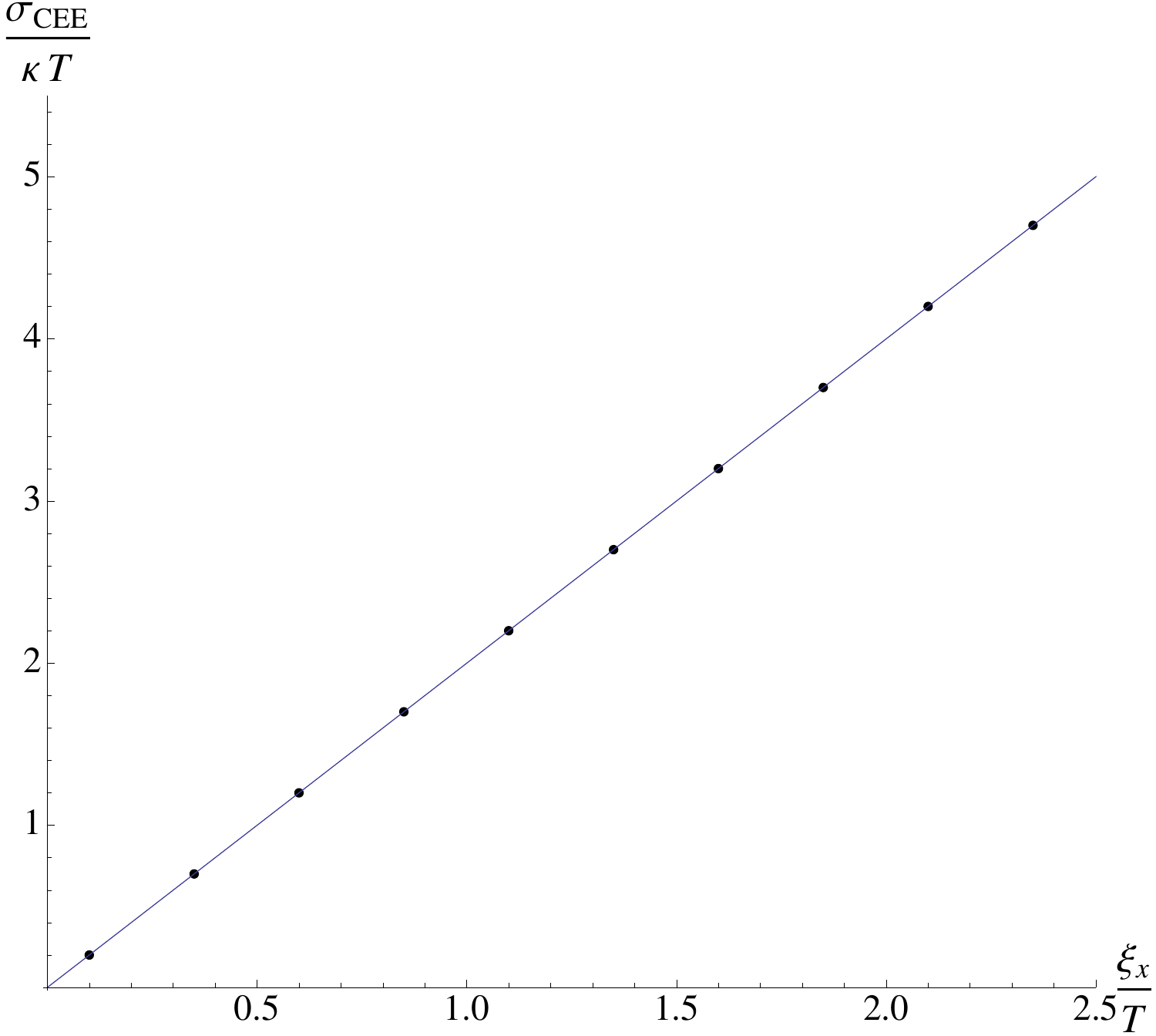}
\caption{\label{fig:CECaxialonlyvsmuandxi} (Left) Chiral electric conductivity versus chemical potential. Each line corresponds to a different superfluid velocity, $\xi_x/T = 0.1-2.1$. We observe that $\sigma_{CEE}/\kappa T=0$ at $\mu_{5c}$ and it approaches a constant value at low temperatures/ large chemical potential. (Right) Dots correspond to $\sigma_{CEE}/\kappa T$ versus $\bar{\xi}_x$ in the large $\mu_5$ region in which $\sigma_{CEE}/\kappa T$ is independent of $\mu_5$. The solid line corresponds to a linear fit; the slope is 2.667. The conductivity depends linearly with $\kappa$. }
\end{figure}
\begin{figure}[h] 
\centering
\includegraphics[width=230pt]{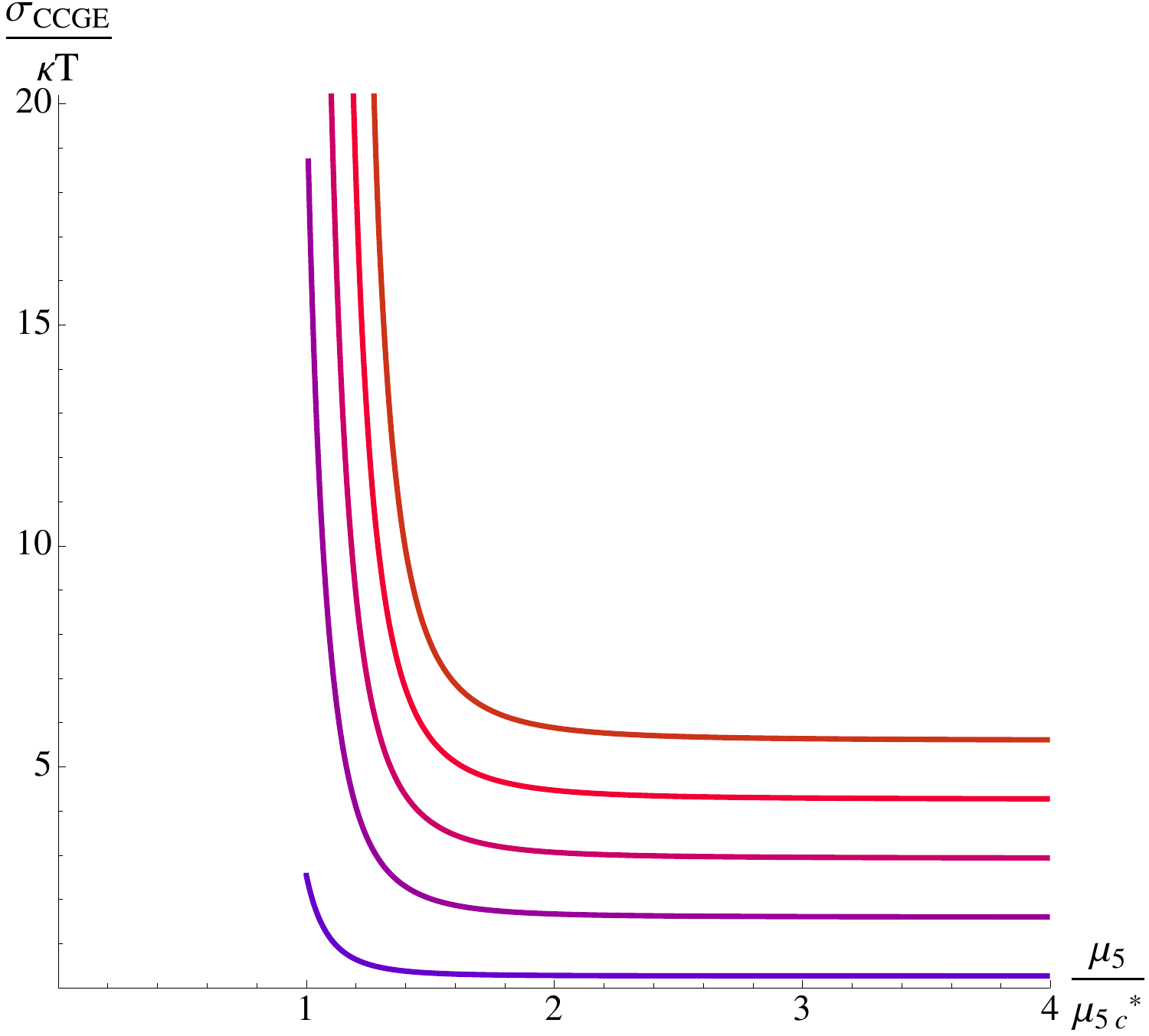}\hspace{2mm}
\includegraphics[width=230pt]{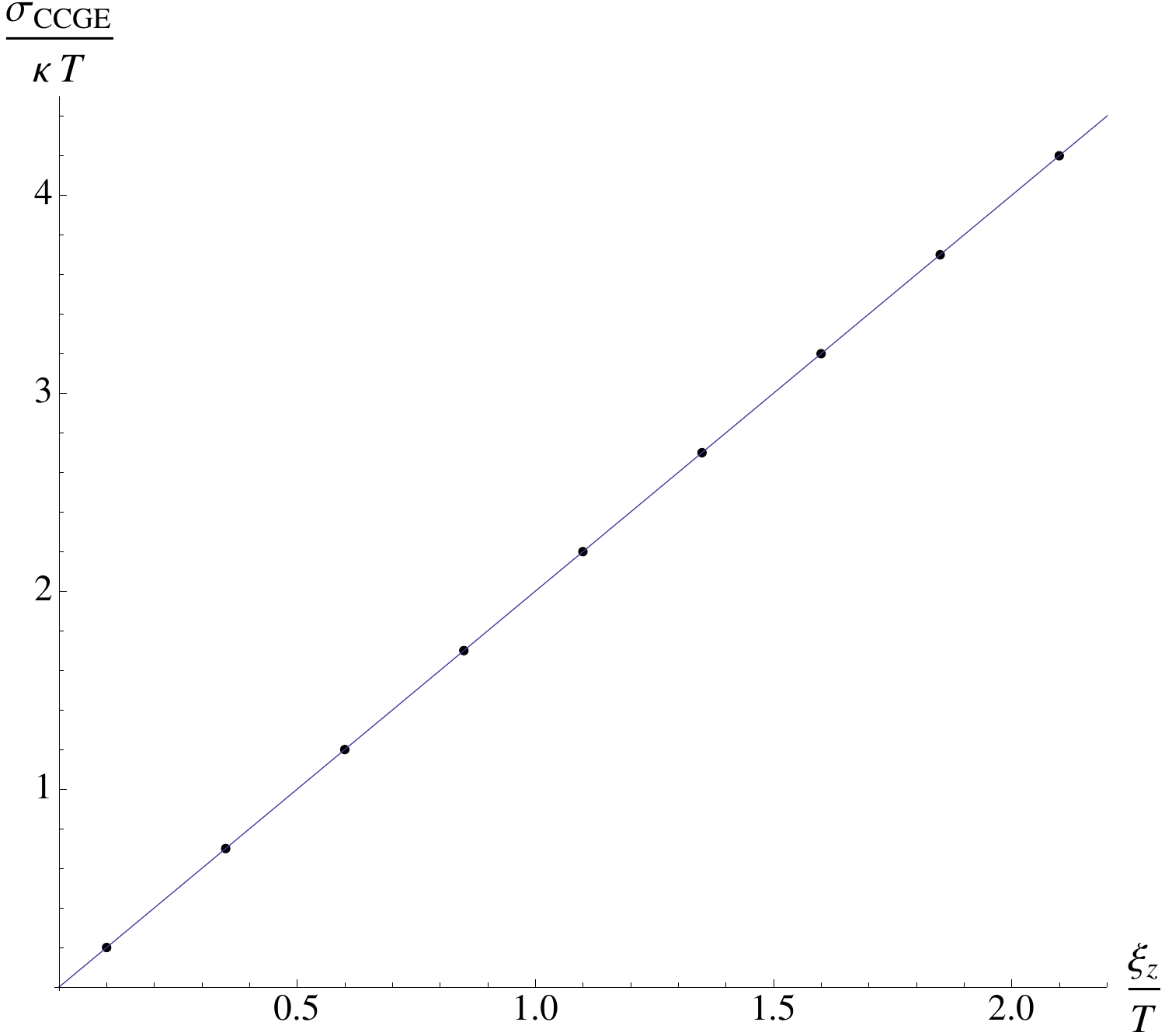}
\caption{\label{fig:CGGEvsmu5} (Left) Chiral charge generation conductivity versus chemical potential, different lines correspond to different values of the supervelocity, $\xi_x/T=0.1-2.1$. (Right) Dots correspond to  $\sigma_{CCGE}$ versus supervelocity for large values of the chemical potential. The solid line corresponds to a linear fit; the slope is 2.667. The conductivity depends linearly with $\kappa$.}
\end{figure}
In order to compute the chiral conductivities from the Kubo formulae (\ref{Chiral1})-(\ref{Chiral 3}) we study perturbations on top of the background we have built. We first want to explore the axial conductivity\footnote{In the literature this conductivity has often been directly associated to the CMC, for the qualitative dependence of the three conductivities of (\ref{Chiral1}) on the axial/vector chemical potentials is the same in the absence of condensate. However, there are significant differences when a condensate distinguishing between axial and vector currents is present, as we will see. Thus, we will stick to the notation of \cite{Gynther:2010ed} and denote as CMC the conductivity related to a vector-vector correlator when a AVV anomaly is switched on.} and the CEC, therefore we switch on the perturbations with non-vanishing frequency and momentum pointing in the direction parallel to the supervelocity (that we choose to be the $x$-direction). The sector we are interested in decouples from the rest of the field perturbations in this kinematic setup, leaving us with just the perturbations of the transverse gauge fields 
\begin{align}
\delta A_y = a_y (r,t,x); \hspace{2cm} \delta A_z = a_z (r,t,x) 
\end{align}
In momentum space the equations read
\begin{align}
\label{eqnpert1axial} a''_y + \left(\frac{f'}{f} + \frac{1}{r}\right) a_y' + \frac{1}{f}\left(\frac{\omega^2}{f} - \frac{k^2 L^2}{r^2} - 2\psi^2\right)a_y + 16 ik \frac{\kappa L }{rf}\phi' a_z + 16 i \omega \frac{\kappa L }{rf}V' a_z=0 \\ 
\label{eqnpert2axial} a''_z + \left(\frac{f'}{f} + \frac{1}{r}\right) a_z' + \frac{1}{f}\left(\frac{\omega^2}{f} - \frac{k^2 L^2}{r^2} - 2\psi^2\right)a_z - 16 ik \frac{\kappa L }{rf}\phi' a_y - 16 i \omega \frac{\kappa L }{rf}V' a_y=0
\end{align}
In the unbroken phase it is possible to find an analytic solution to the above system of equations in the kinematic limit $\omega=0$ and to first order in momentum $k_x \equiv k$. Recall that this is all that we need in order to obtain $\sigma_{55}$, making use of Kubo formulae \cite{Amado:2011zx}. However, in the case at hand the background has been computed numerically and therefore we will look directly for numerical solutions to the system (\ref{eqnpert1axial})-(\ref{eqnpert2axial}).\\ % henceforth we will be using data for a background with $m^2=-7/2$.\\
We are now ready to calculate the Kubo formulae shown in (\ref{Chiral1})-(\ref{Chiral 3}). The problem reduces to numerically computing the two retarded 2-point functions with the usual holographic prescription \cite{Kaminski:2009dh} (see the Appendix for details on the computation).

A comment that applies to all figures is in order here. The critical value of the chemical potential depends on the value of the supervelocity and the mass of the scalar field. In our convention, $\mu^*_c$ is the critical value at zero supervelocity and $m^2=-7/2$.\\

Our results for $\sigma_{55}$ are depicted in Figure \ref{fig:Sigma55vsV}. We observe that $\sigma_{55}$ is proportional to the (axial) chemical potential even in the broken phase. However, the coefficient of proportionality decreases from 1 to 1/3 in units of $e^2 N_c/4 \pi^2$. Numerically, in terms of $\kappa$ we get\footnote{In order to make contact with the computation in the unbroken phase of  \cite{Amado:2011zx}, notice that we have set $16\pi G\equiv 1$ in (\ref{action}). Hence, their result $\sigma^{\text{unbrok.}}_B = 8 \kappa \mu_5/(16\pi G)$ corresponds to $\sigma^{\text{unbrok.}}_{55} = 8  \mu_5 \kappa$ with our conventions.}
\begin{align}
\frac{\sigma_{55}\left(\frac{\bar \mu_5}{\bar \mu_{5c}} >>1\right) }{\kappa \mu_5}=2.668 \approx \frac{8}{3}\,.
\end{align} 
This reduction has been predicted to be universal. In our model, we can check that this is independent from the mass of the bulk scalar field (right plot of Figure \ref{fig:Sigma55vsV}). Remarkably, finite supervelocity does not alter these conclusions, as depicted in Figure \ref{fig:Sigma55vsV} (left); the correction to the transport coefficient is independent of the supervelocity. As a final remark, we find that the dependence of the axial conductivity with $\kappa$ is unaffected by the presence of the condensate and the supervelocity, namely $\sigma_{55}\sim\kappa$.\\
Moving to the CEC, we observe that it starts increasing but rapidly approaches a constant value, independent of $\mu_{5}/T$. On the contrary, it linearly increases with the superfluid velocity for large chemical potential, see Figure \ref{fig:CECaxialonlyvsmuandxi}. Or results thus strongly suggest that, at low temperatures,
\begin{align}
\frac{\sigma_{CEE}\left(\frac{\bar \mu_5}{\bar \mu_{5c}} >>1\right)}{\kappa T} =  2.667 \frac{\xi_x}{T} \approx\frac{8}{3}\frac{\xi_x}{T} \,.
\end{align}
Notice that this value is essentially the same as the observed for $\sigma_{55}$ at large axial chemical potential. Again the dependence with $\kappa$ is linear.
\subsection{The Chiral Charge Generation Effect}
\label{subsec:CCGEax}
Let us now induce a supervelocity in the $z$-direction, by turning on $A_z(r)$ instead of $A_x(r)$ in the bulk. This, as anticipated, influences the quasinormal modes, even though the background equations remain the same as in the previous subsection (due to the fact that, without superflow, the background is isotropic), with the replacement $A_x \leftrightarrow A_z$. On top of this we switch on perturbations with non-vanishing frequency and momentum pointing in the $x$-direction (transverse to the supervelocity). The equations for the perturbations in the transverse sector are more involved now for they couple to all other perturbations. They can be found in Appendix \ref{primerastransverse} .\\
As mentioned in the introduction, the CCGE corresponds to a "generation" of charge proportional to the scalar product of the supervelocity and the magnetic field
\begin{align}
\rho =\hat \sigma\  \vec \xi \cdot  \vec B\,.
\end{align}
As aforementioned, for convenience we will absorb the supervelocity component into the conductivity, i.e. $\sigma_{CCGE}  =\hat \sigma \xi_z$. Note that the charge vanishes if the supervelocity is parallel to the external momentum. In order to observe such an effect, we will use (\ref{CCGC}).\\
We proceed as before and present our result in Figure \ref{fig:CGGEvsmu5} . We observe that indeed this phenomenon is not negligible in the presence of supervelocity. Moreover, it stabilizes at large enough chemical potential; in the region in which $\sigma_{CCGE}$ does not depend on $\bar \mu_5$, it presents a clear linear dependence on the superfluid velocity (right plot of Figure \ref{fig:CGGEvsmu5}). We can perform a numerical quadratic fit, obtaining
\begin{align}
\frac{\sigma_{CCGE}\left(\frac{\bar \mu_5}{\bar \mu_{5c}} >>1\right)}{\kappa T} =2.667\frac{\xi_z}{T} \approx\frac{ 8}{3}\frac{\xi_z}{T}
\end{align}
to a good approximation. Again, the slope has the same value as for the CEC. Let us emphasize that the behaviour of this transport coefficient at the phase transition is strange at first sight. Naively, we would have expected $\sigma_{CCGE}(\bar \mu_c) =0$ instead of the observed value. We comment on this issue in Section \ref{sec:conc}.

\section{Model with axial and vector currents}
\label{sec:axialvec}
\begin{figure}[h] 
\centering
\includegraphics[width=230pt]{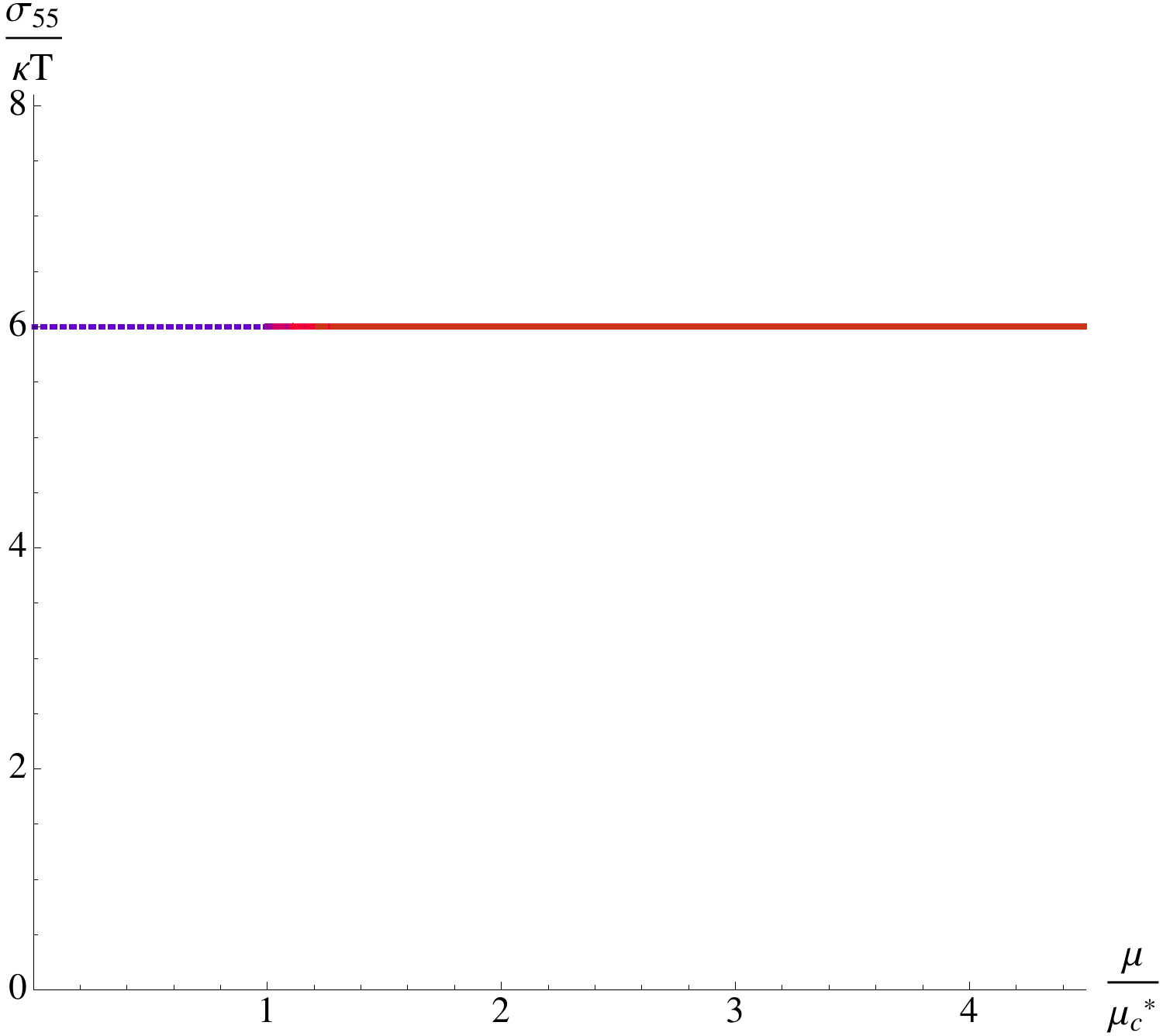}\hspace{2mm}
\includegraphics[width=230pt]{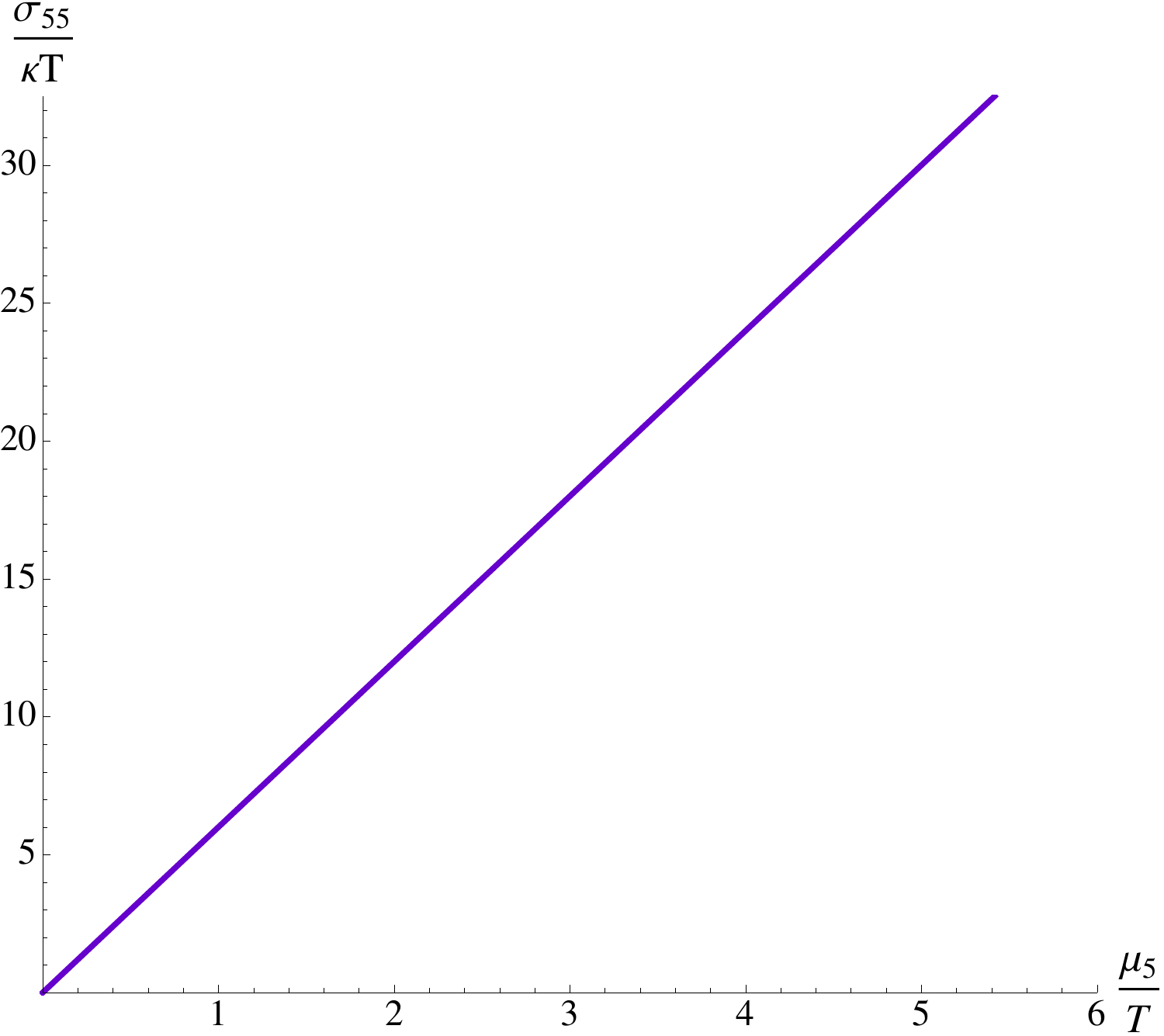}
\caption{\label{fig:sigma55} (Left) Axial conductivity versus vector chemical potential at $\bar \mu_5=1$ and $\xi_x/T= 0.1-2.1$. We find that $\sigma_{55}$ is independent of both the vector chemical potential and the superfluid velocity. (Right) $\sigma_{55}$ versus axial chemical potential. The dependence with $\mu_5$  is linear, as expected. The conductivity depends linearly with $\kappa$.}
\end{figure}

\begin{figure}[h] 
\centering
\includegraphics[width=230pt]{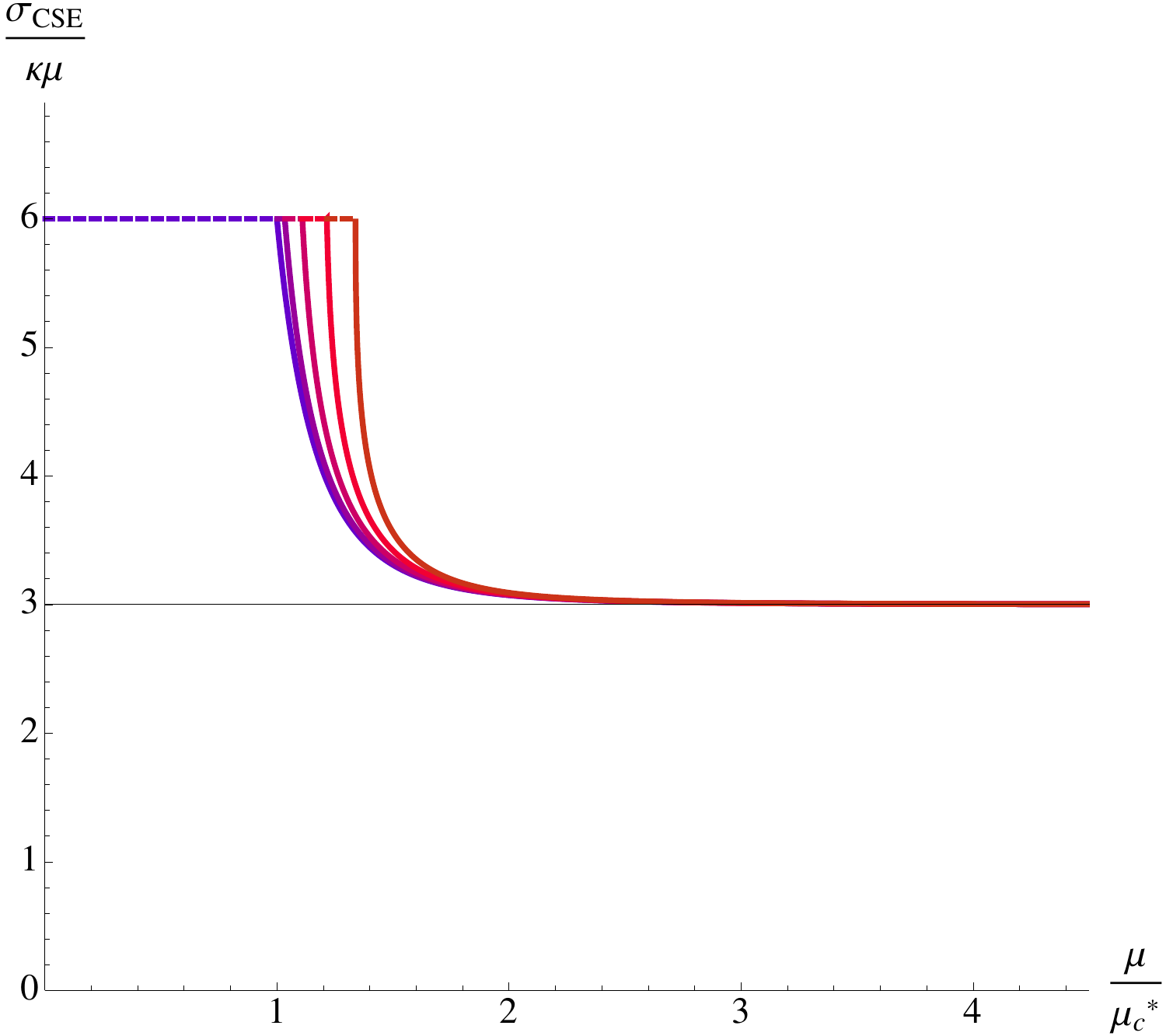}\hspace{2mm}
\includegraphics[width=230pt]{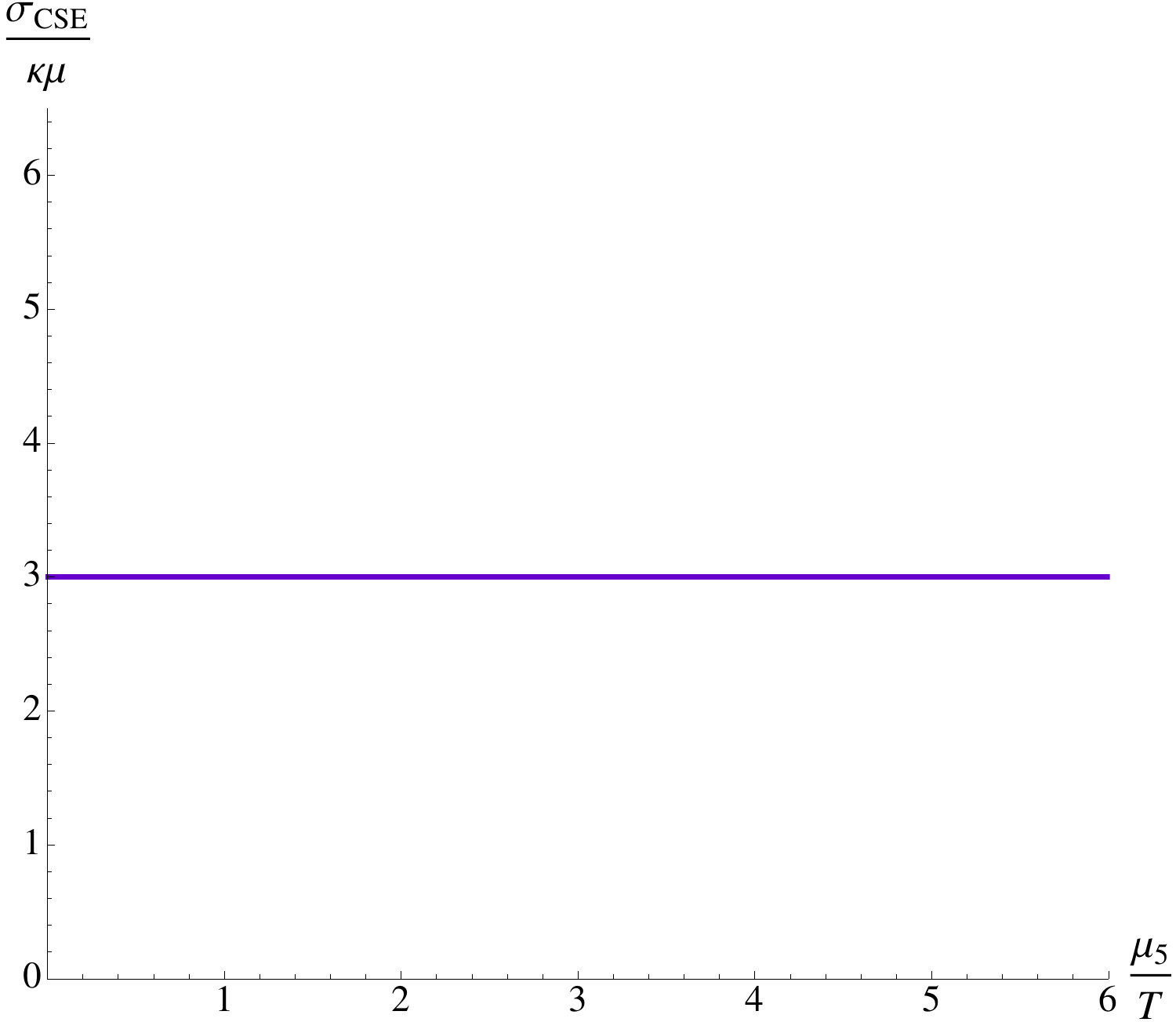}
\caption{\label{fig:sigmaCSE} (Left) Chiral separation conductivity divided by vector chemical potential versus vector chemical potential, $\bar \mu_5=1$ and $\xi_x/T= 0.1-2.1$. The conductivity now approaches $1/2$ of the value at $\bar \mu_c$, independently of $\xi_x/T$. (Right) The plot shows this conductivity against the axial chemical potential for generic values of $\mu$. $\sigma_{CSE}$ is independent of the axial chemical potential in both the broken and ubroken phases. The conductivity depends linearly with $\kappa$.}
\end{figure}

\begin{figure}[h] 
\centering
\includegraphics[width=230pt]{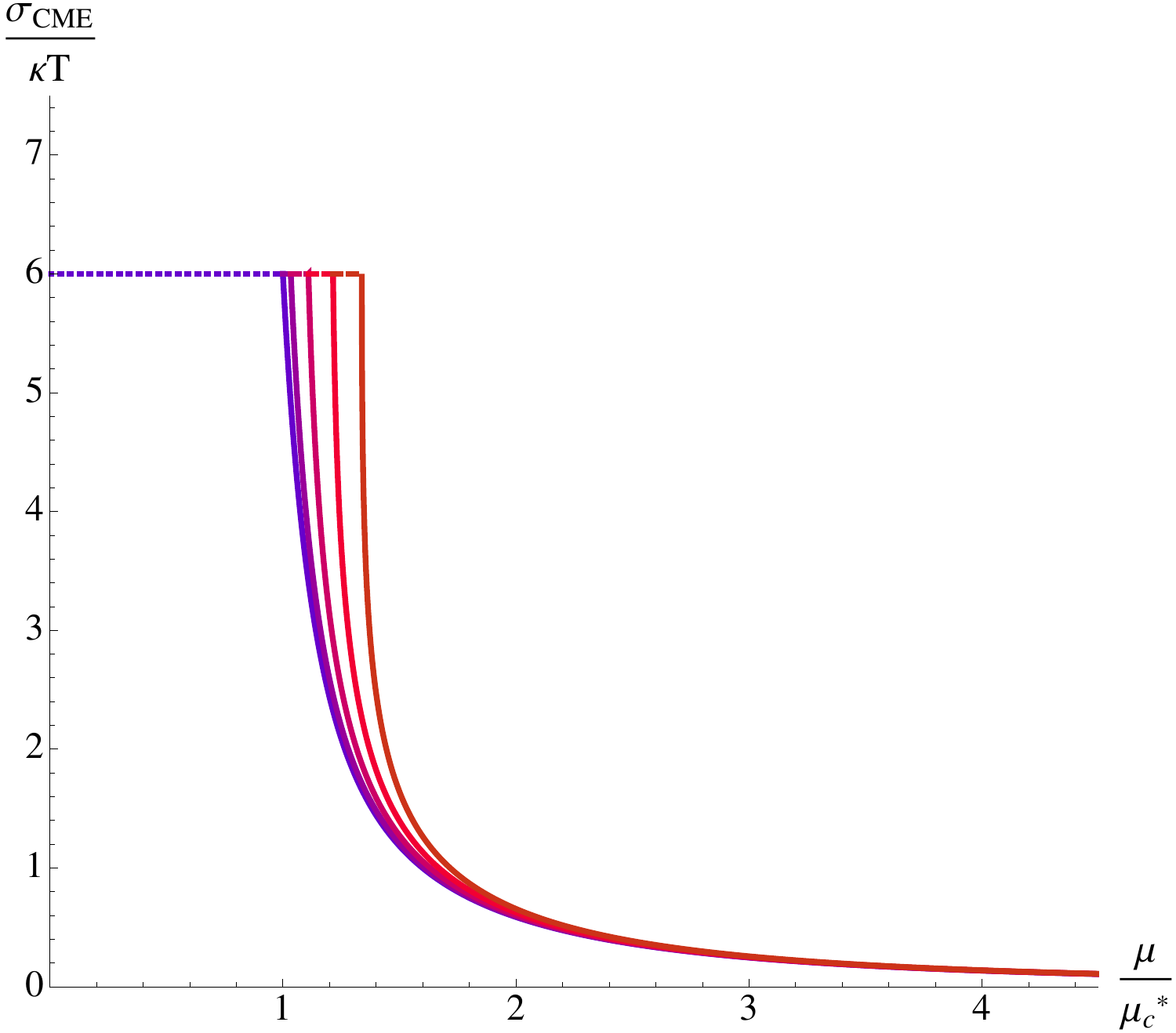}\hspace{2mm}
\includegraphics[width=230pt]{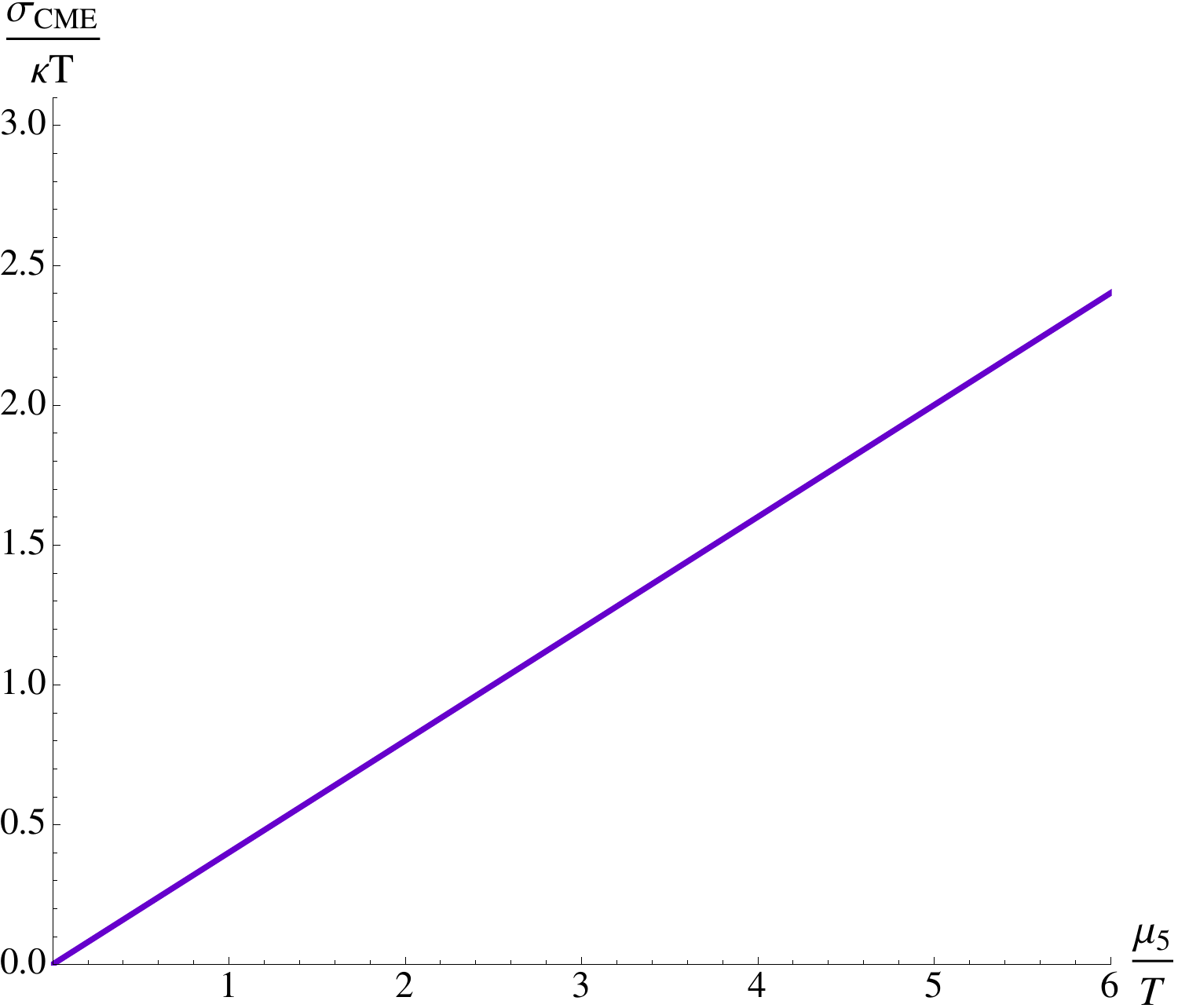}
\caption{\label{fig:sigmaaxCME} (Left) Chiral magnetic conductivity versus vector chemical potential with $\bar \mu_5=1$. Different lines correspond to different values of the superfluid velocity, with $\xi_x/T=0.1-2.1$. The best fit shows that for large enough values of $\bar \mu$ it decreases as $\sigma \sim 1/\bar \mu^2$. (Right) $\sigma_{CME}/ \kappa T$ vs. axial chemical potential with $\mu/ T= 2.5$. The linear dependence with $\mu_5$, characteristic of the unbroken phase, remains unaltered. The conductivity depends linearly with $\kappa$.}
\end{figure}

\begin{figure}[h] 
\centering
\includegraphics[width=230pt]{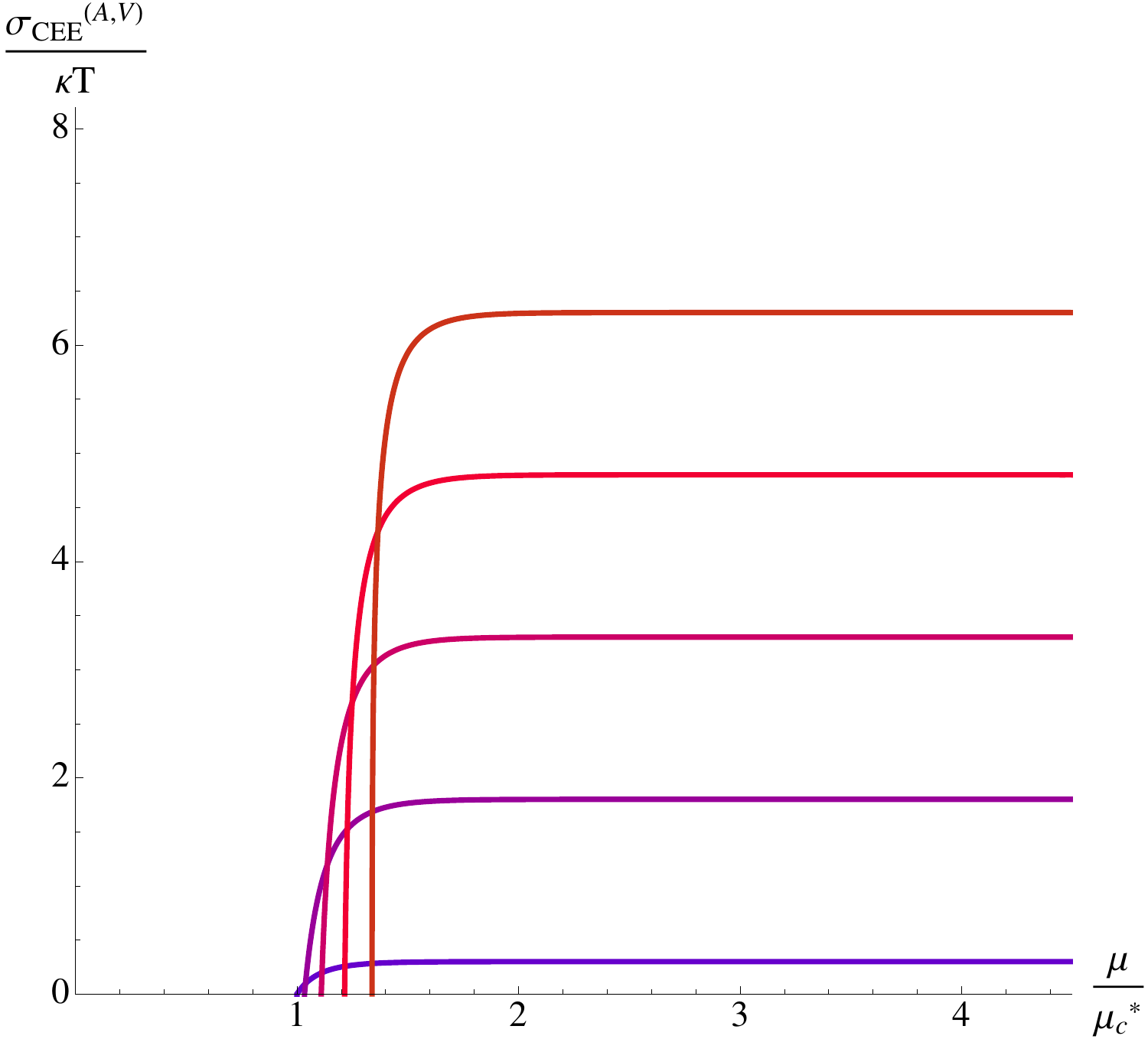}\hspace{2mm}
\includegraphics[width=230pt]{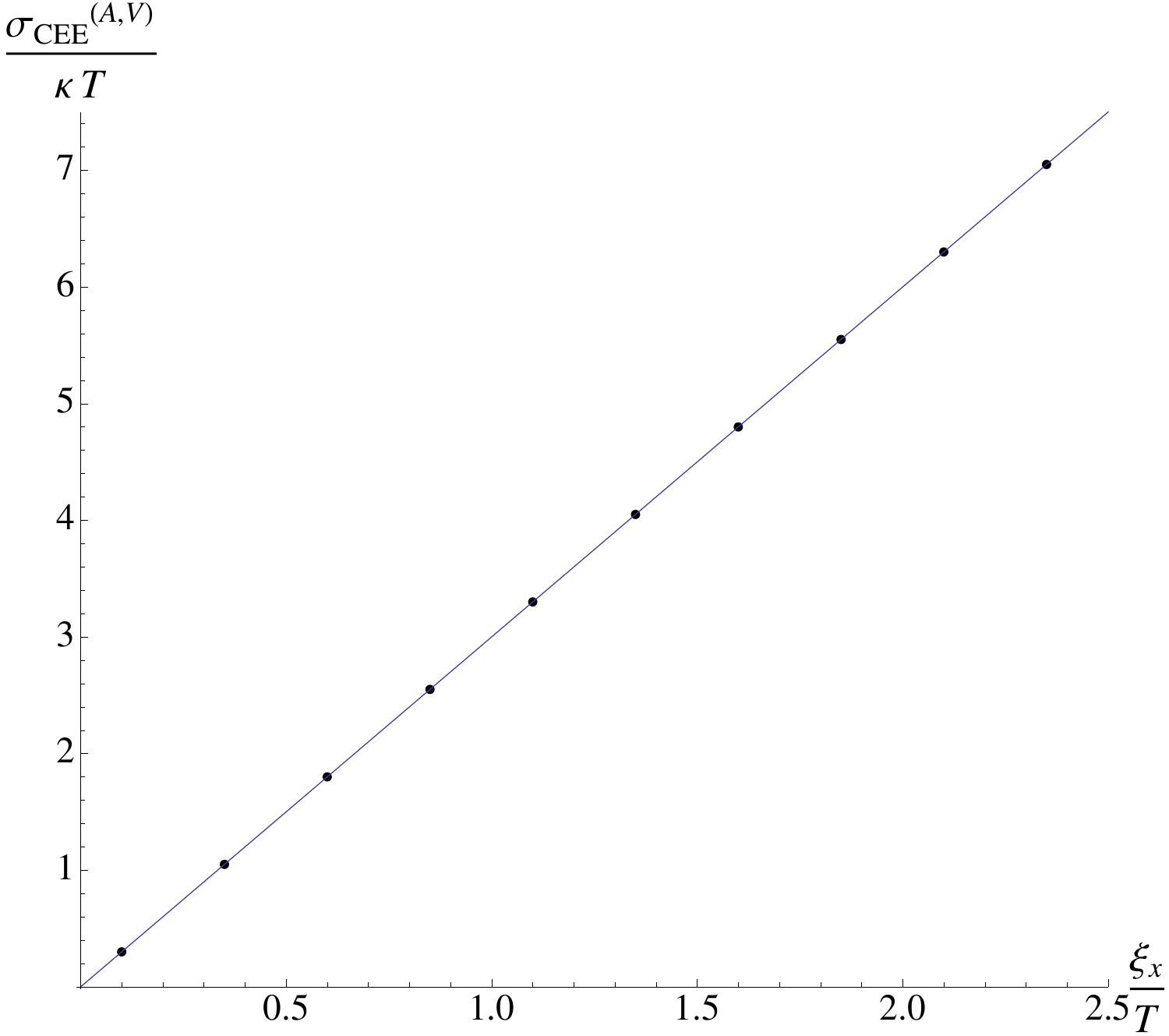}
\caption{\label{fig:CECaxialvec} (Left) Chiral electric conductivity versus vector chemical potential at  $\bar \mu_5= 1$ and $\xi_x/T= 0.1-2.1$ (bottom to top). Both $\sigma^{V}_{CEE} /T$ and $\sigma^{A}_{CEE} /T$ show the same behaviour . (Right) Chiral electric conductivities versus supervelocity at  $\bar \mu_5=1$ in the region where they don't depend on $\mu$. The conductivity depends linearly with $\kappa$.}
\end{figure}
In this section we study the more realistic model, in which we consider two $U(1)$ bulk gauge fields, being only one of them spontaneously broken. There are two different interpretations of this model:
\begin{itemize}
\item We have axial and vector currents $U(1)_V \times U(1)_A$ and the condensate is coupled only to the vector part, whereas the axial symmetry is unbroken. This realizes the interplay between anomalous axial and vector currents, first considered in \cite{Gynther:2010ed}. \\ The fact that the axial current is not coupled to the scalar field means that the axial charge of the condensate is zero, so the axial chemical potential can be made constant through the phase transition and is not affected by the condensation whatsoever.  
\item The unbroken $U(1)$ is a generic field and the two $U(1)$'s are intertwined in a particular way by the anomaly. With this second interpretation, crossed anomalous correlators can be related to the response of the (broken) current to an external unscreened magnetic field, associated to the unbroken symmetry. This avoids any possible problem with the physical realization external magnetic fields contained in the bulk of the system.
\end{itemize}
Despite of the two possible interpretations, we will use a notation adapted to the first one. The action of the model contains a complex scalar field coupled to the vector sector
\begin{align}
\label{Lagaxvec} \mathcal{L}=-\frac{1}{4} F_{MN}F^{MN} -\frac{1}{4} G_{MN}G^{MN} + \frac{\kappa}{2}\epsilon^{MABCD}A_M(3 F_{AB}F_{CD}+G_{AB}G_{CD}) - \overline{D_{M}\Psi} D^M\Psi - m^2 \bar\Psi \Psi \,.
\end{align}
Here $F$ is the field strength for the vector gauge field $V$ and $G$ is the analogue for the axial gauge field $A$. Moreover $D_{M}\Psi = \partial_M \Psi - i V_M \Psi$. We consider AAA and AVV anomalies. \\
The equations of motion for the background are the same as (\ref{backg1})-(\ref{backg3}), with an additional equation for the background axial gauge field $A(r)= K(r) dt$
\begin{align}
\label{backax} K'' + \frac{3}{r} K' =0
\end{align}
which has a trivial analytic solution $K(r) = K_0 - K_1/r^2$. The boundary conditions for the gauge fields are:
\begin{align} 
\phi(r)_{r\rightarrow \infty}\sim \mu \hspace{2cm} V(r)_{r\rightarrow \infty}\sim \xi_{\{x,z\}}  \hspace{2cm} K(r)_{r \rightarrow \infty}\sim  \mu_5
\end{align}
We impose again standard quantization for the scalar field. First we choose the supervelocity to point in the $x$-direction.
On top of this we switch on the perturbations with non-vanishing frequency and momentum parallel to the supervelocity. The equations for the perturbations in the transverse sector can be found in Appendix \ref{eqsparalel}. There is a wider set of correlators that we can study in this set up
\begin{align}
\label{CSC} \sigma_{55}= \lim_{k\rightarrow 0} \frac{i}{2k} \left<J_A^y J_A^z\right>_{\mathcal{R}}(\omega=0, k)\\
\label{CAC} \sigma_{CSE}= \lim_{k\rightarrow 0} \frac{i}{2k} \left<J_V^y J_A^z\right>_{\mathcal{R}}(\omega=0, k)\\
\label{CMCV} \sigma_{CME} = \lim_{k\rightarrow 0} \frac{i}{2k} \left<J_V^y J_V^z\right>_{\mathcal{R}}(\omega=0, k)
\end{align}
In the superfluid phase, after assuming that the supervelocity is transverse to the momentum, we can also consider the Kubo formulae related to the Chiral Electric Effect and the Chiral Charge Generation Effect
\begin{align}
\label{CCGCV} &\sigma^{A}_{CCGE} = \lim_{k\rightarrow 0} \frac{i}{2k_{\bot}} \left<J^0_{A} J^y_{V}\right>_{\mathcal{R}}(\omega=0, k)\ ;\hspace{1.5cm} \sigma^{V}_{CCGE} = \lim_{k\rightarrow 0} \frac{i}{2k_{\bot}} \left<J^0_{V} J^y_{A}\right>_{\mathcal{R}}(\omega=0, k)\\
\label{CECV}&\,\,\,\sigma^{A}_{CEE} = \lim_{k\rightarrow 0} \frac{i}{2\omega} \left<J^y_{A} J^z_{V}\right>_{\mathcal{R}}(\omega, k=0)\ ; \hspace{1.95cm}\sigma^{V}_{CEE}= \lim_{k\rightarrow 0} \frac{i}{2\omega} \left<J^y_{V} J^z_{A}\right>_{\mathcal{R}}(\omega, k=0)
\end{align}
We expect them to receive different corrections due to the fact that the condensate distinguishes between the vector and the axial symmetry. Notice that our notation establishes that, for example, $\rho_A = \sigma^{A}_{CCGE}B^V_{z}$ and $\rho_V = \sigma^{V}_{CCGE} B^A_z$.\\

Our results are as follows. On the one hand, the correlator $\left<J^y_5 J^z_5\right>$ does not get altered due to the condensate, and is linear in $\mu_5$, as depicted in Figure \ref{fig:sigma55}. The behavior could have been anticipated, since the on-shell action is diagonal in vector/axial sectors and it is clear that in the dynamical equations (\ref{eqnpert1broax})-(\ref{eqnpert2broax}) the mixing between $a_y$ and $a_z$ is independent of the condensate. This is ultimately due to the fact that the condensate only couples to the vector sector and that the correlator $\left<J^y_5 J^z_5\right>$ is only sensitive to the AAA anomaly\footnote{The independence of the condensate can be spoiled by altering the model. For instance, by inducing an axial component for the condensate.}.

On the other hand, the results concerning $\sigma_{CSE}$ are summarized in Figure \ref{fig:sigmaCSE}. This conductivity acts similarly to that encountered in the first section. This was expected by the form of the equations of motion: in this model, the correlator mixing between $a_y$ and $v_z$ is mediated by the same background fields as in the model with only axial symmetry. Remarkably, unlike the case with a $U(1)^3$ anomaly, at large values of $\bar \mu$ we obtain\footnote{The numerical value $\sigma_{CSE}(T_c)/(\kappa \mu) \approx 6$ depends on the strength of the $\kappa$-term in the equations of motion and is not of fundamental importance, for it can be easily rescaled (compare to Section \ref{sec:Brokaxial}).}
\begin{align}
\label{CSET0} \frac{\sigma_{CSE} \left(\frac{\bar \mu}{\bar \mu_{c}} >>1\right) }{\kappa\mu}= 2.998 \approx \frac{\sigma_{CSE}(\bar \mu_c)}{2 \kappa \mu} \,,
\end{align}
independently of the superfluid velocity. This result indicates that the $T\rightarrow0$ behaviour is strongly dependent on the structure of the broken symmetries and the interplay of the anomalies. Moreover, the conductivity does not depend on the axial chemical potential (right plot).

Finally, let us comment on the $\sigma_{CME}$. The results are displayed in Figure \ref{fig:sigmaaxCME}. We find a linear dependence on $\bar \mu_5$, as expected. However, in the presence of the condensate we observe a new dependence on the vector chemical potential, which is absent in the unbroken phase. The chemical potential diminishes the value of the CMC strongly and it tends to zero for large values of $\bar \mu$ as
\begin{align}
\frac{\sigma_{CME}\left(\frac{\bar \mu}{\bar \mu_{c}} >>1\right)}{\kappa T} \approx g \frac{1}{\bar \mu^2}
\end{align}

with a numerical value for $g \approx 2.15$. We elaborate on this in Section \ref{sec:conc}. %This result fits more with the spirit of the behaviour of $\sigma_{CME}$ observed when entering a (chiral) phase transition, both by lattice simulations [REF.] and analytic computations [REF.].

For the chiral transport coefficients associated to the CEE, we observe that correlators of the form $\left<J_A J_A\right>(k=0)$ and $\left<J_V J_V\right>(k=0)$ vanish identically. Concerning the ones mixing axial and vector currents, we observe that $\sigma^V_{CEE}=\sigma^A_{CEE}\equiv \sigma^{(V,A)}_{CEE}$. The result is depicted in Figure \ref{fig:CECaxialvec}. Fitting the right plot to a parabola, we get 
\begin{align}
\label{CECAVT0} \frac{\sigma^{(V,A)}_{CEE} \left(\frac{\bar \mu}{\bar \mu_{c}} >>1\right)}{\kappa T}= 3.003  \frac{\xi_x}{T}.
\end{align}
with remarkable precision.

\subsection{$U(1) \times U(1)$ model with transverse supervelocity}
\begin{figure}[h] 
\centering
\includegraphics[width=220pt]{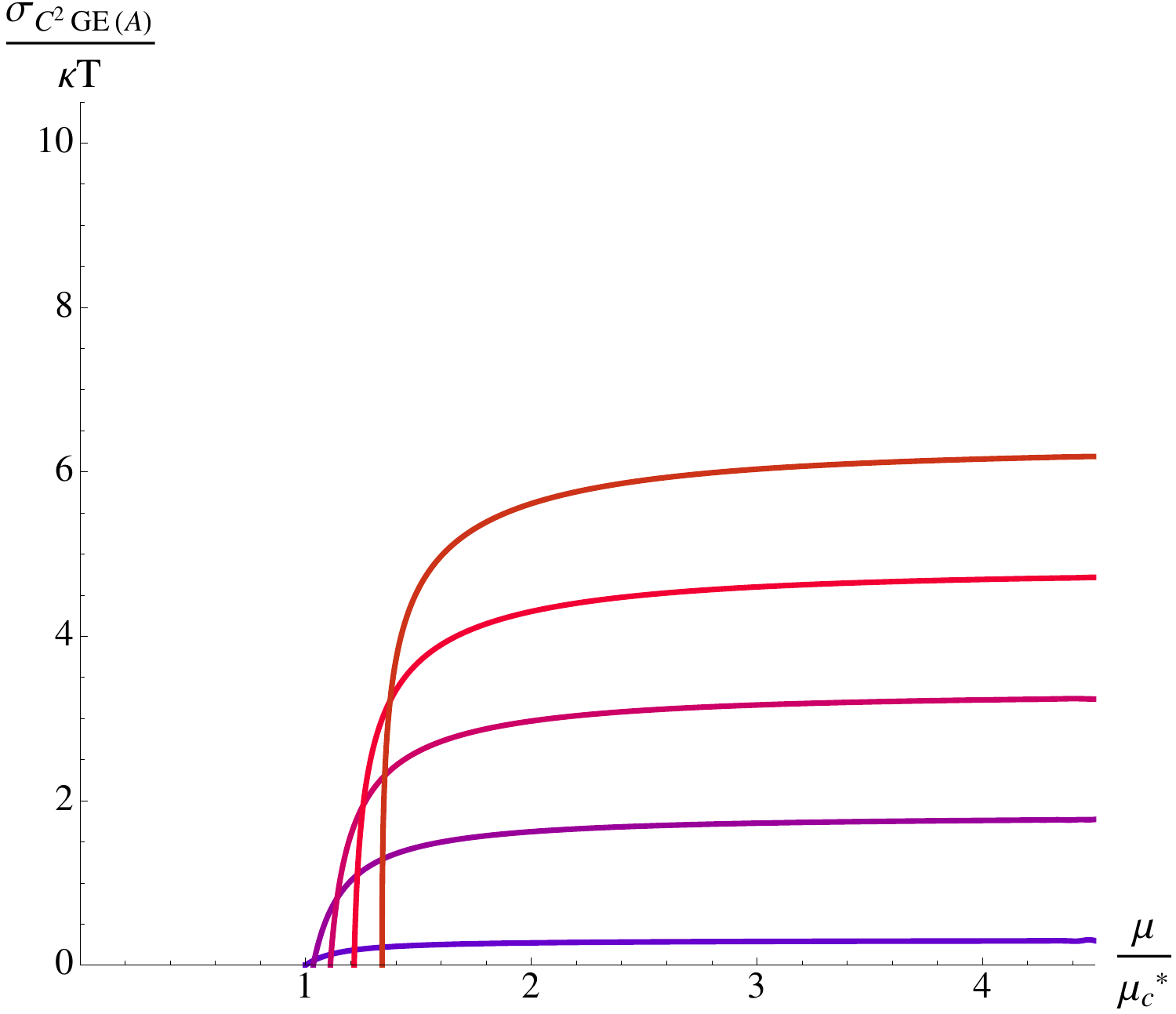}\hspace{2mm}
\includegraphics[width=220pt]{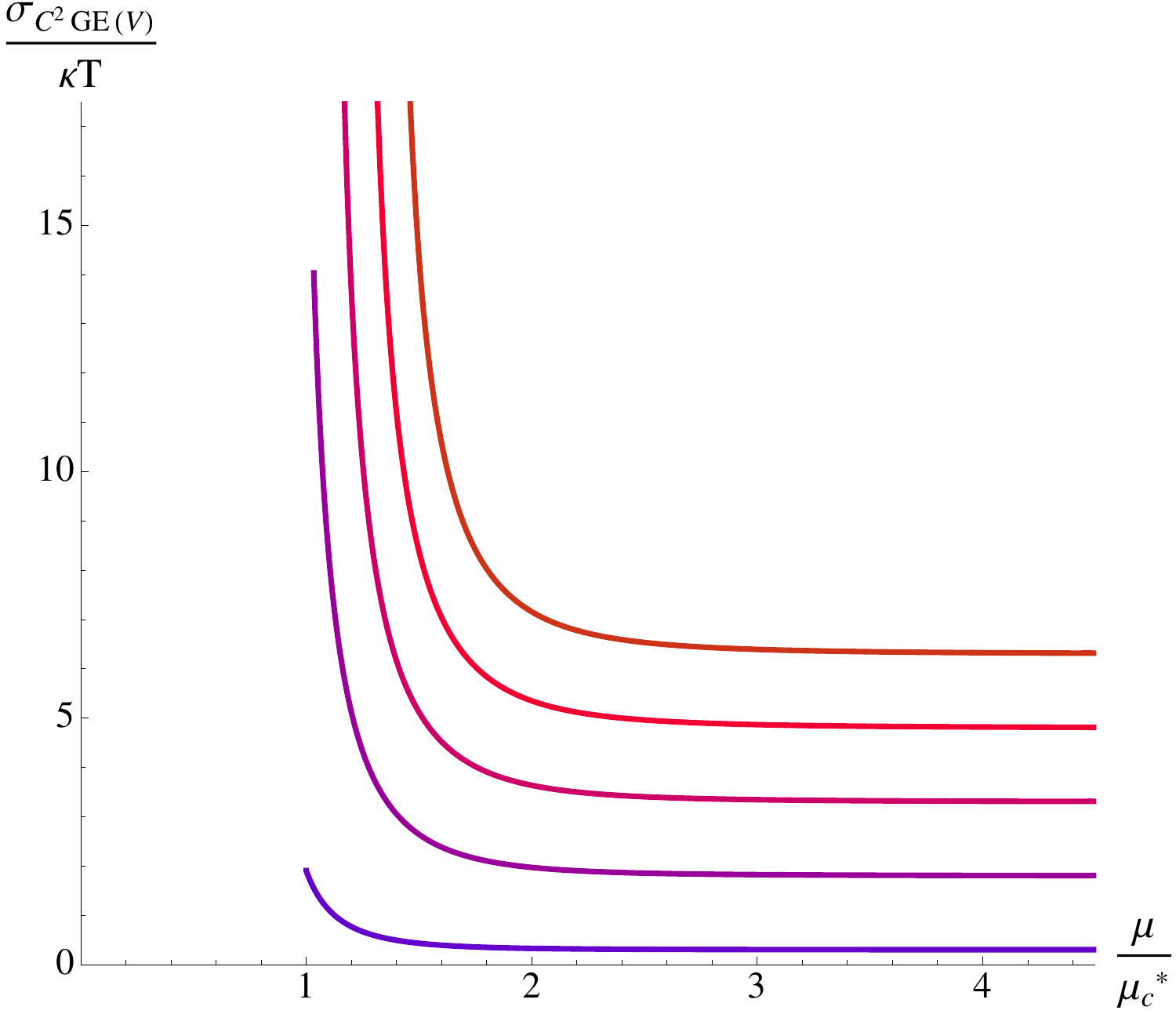}
\caption{\label{fig:CCGA} $\sigma^{(A)}_{CCGE} /\kappa T$ (Left) and $\sigma^{(V)}_{CCGE} / \kappa T$ (Right) versus vector chemical potential at  $\bar \mu_5=1$ and $\xi_x/T= 0.1-2.1$ (bottom to top). For large enough values of the chemical potential both conductivities show the same behaviour. Both depend linearly with $\kappa$}
\end{figure}

\begin{figure}[h] 
\centering
\includegraphics[width=220pt]{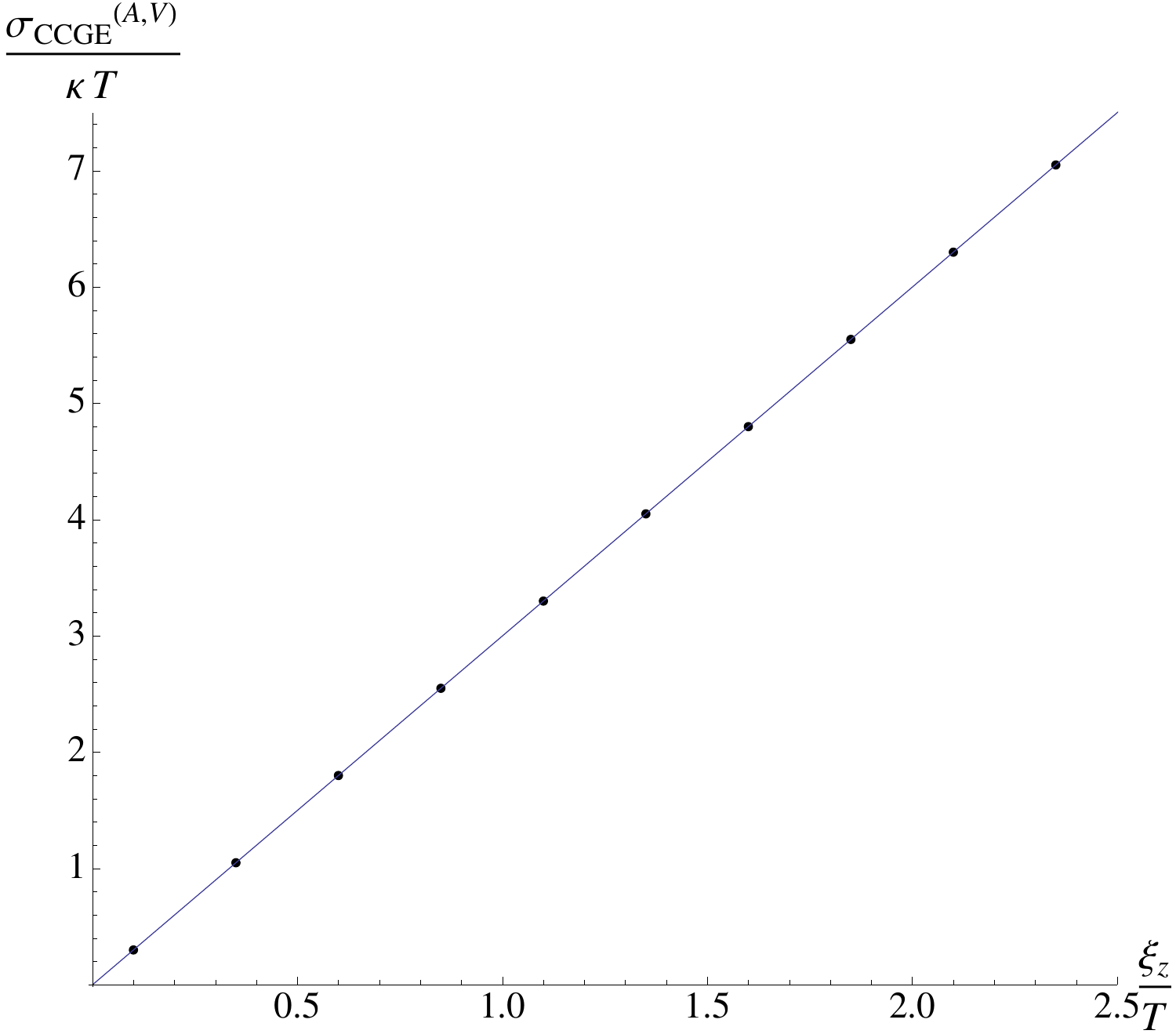}
\caption{\label{fig:CCGV} Both conductivities $\sigma^{(A,V)}_{CCGE}$ show the same dependence on the vector chemical potential $\mu$ and the supervelocity $\xi_z$ for large enough values of $\mu$. The slope coincides with the slope for the CEC, despite the radically different behaviour close to the phase transition.}
\end{figure}

As we did in the previous model, in order to study the CCGE we switch on perturbations with non-zero frequency and momentum pointing in the $x$-direction, transverse to the superfluid velocity ($z$-direction). The system of equations with transverse supervelocity can be found in Appendix \ref{ultimapendix}. We report the results on the CCGC in Figure \ref{fig:CCGA}. \\
As shown in there are now two different conductivities related to the CCGE, which we denote $\sigma^{(V)}_{CCGE}$ and $\sigma^{(A)}_{CCGE}$   They exhibit a very different behavior close to $\bar \mu_c$; the conductivity $\sigma^{(V)}_{CCGE}$ is similar to the one found in Section \ref{subsec:CCGEax}, whereas $\sigma^{(A)}_{CCGE}$ looks like the CEC, with a good continuous behavior at the phase transition. We comment on those differences in Section \ref{sec:conc}. At low temperatures, however, both $\sigma^{(A)}_{CCGE}$ and $\sigma^{(V)}_{CCGE}$ tend to the same value and the dependence with the supervelocity is linear (Figure \ref{fig:CCGV}). A quadratic fit yields
\begin{align}
\frac{\sigma^{(V,A)}_{CCGE} \left(\frac{\bar \mu}{\bar \mu_{c}} >>1\right)}{\kappa T} = 3.003  \frac{\xi_z}{T}.
\end{align}

\begin{figure}[h] 
\centering
\includegraphics[width=220pt]{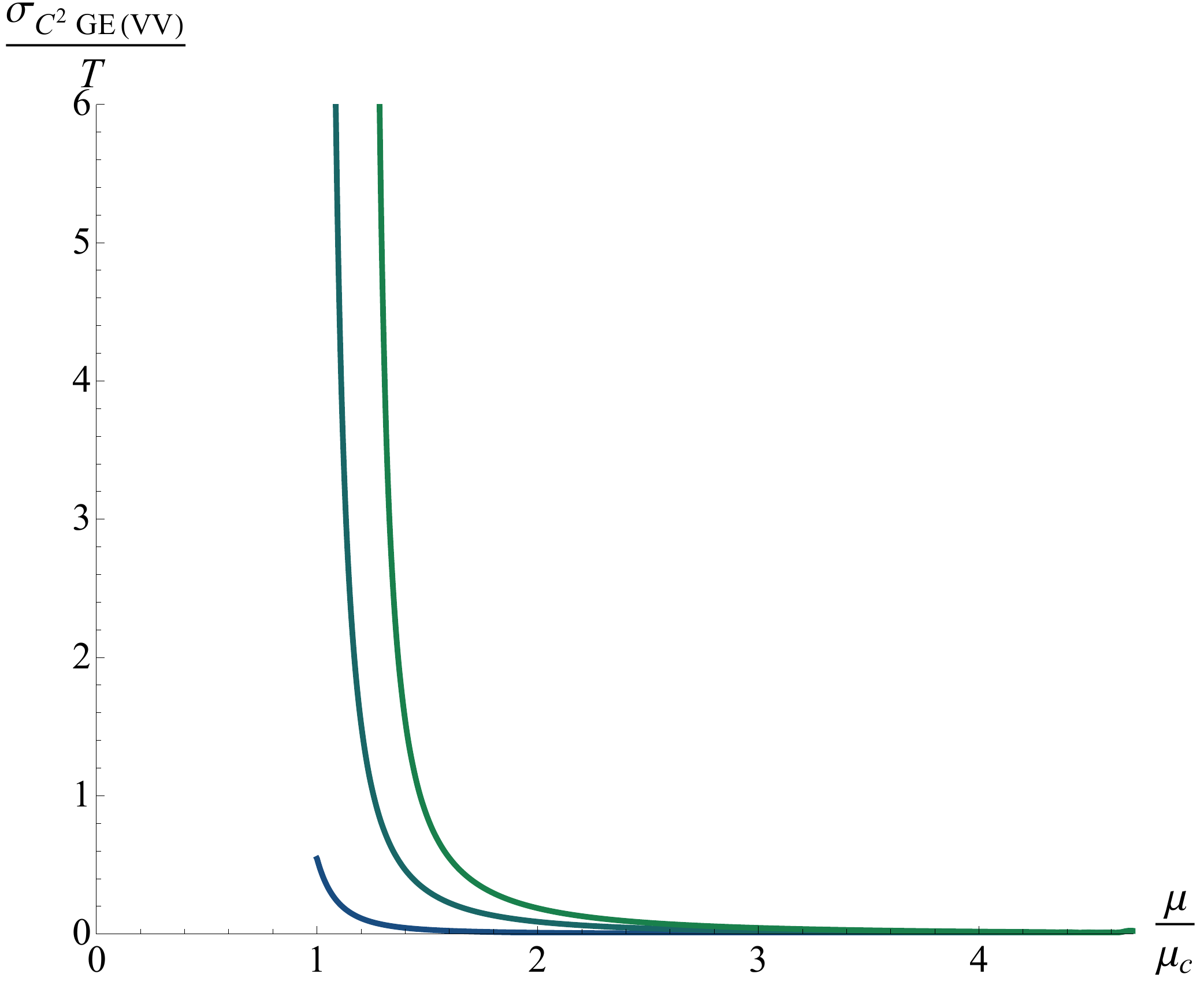}
\caption{\label{fig:rarete} Plot of the $\sigma_{CCGE} ^{(VV)}$ conductivity (defined in the text) versus vector chemical potential for several values of the supervelocity.}
\end{figure}

Remarkably enough, we point out that the conductivity 
\begin{align}
\sigma_{CCGE}^{(VV)} = \lim _{k \rightarrow 0} \frac{i}{2k_{\bot}} \left<J^0_V J^y_V\right>_{\mathcal{R}}(\omega=0, k)\,,
\end{align}
depicted in Figure \ref{fig:rarete}, is not negligible. In principle we could have anticipated it to vanish because of the structure of the anomalies included in the Lagrangian (\ref{Lagaxvec}). As shown in the plot, this only occurs far enough form the phase transition. This effect points towards an "effective VVV anomaly" (see also the results concerning the CMC) that is present close to the phase transition.
\section{Conclusions, Educated guesses and Future directions}
\label{sec:conc}
We have analyzed the explicit form of the chiral conductivities in two holographic models with $U(1)$ and $U(1)\times U(1)$ symmetries, in presence of a scalar condensate, at finite superfluid velocity. 
We have presented an explicit calculation of CEE by using a suitable Kubo formula, which allowed us to prove in a robust way that the CEC is in general not vanishing in superfluids.\\
Moreover, by means of the Kubo formulae we have found an effect whose existence, as far as we are aware, had not been emphasized before in the literature. This induces the presence of axial charge in the presence of supervelocity and a magnetic field
\begin{align}
\rho_A \propto \vec \xi \cdot \vec B
\end{align}
Such a term has interesting consequences. For instance, the Chiral Magnetic Effect would be dynamically produced in a superfluid in the presence of an external magnetic field aligned with the supervelocity. We believe this term deserves more investigation in the future, in order to fully understand the mechanism by which charge is "generated", as well as to analyze the implications that it could lead to.\\
In addition, we have found generic corrections, due to the background condensate, to all of the anomalous conductivities. Such corrections seem to take a constant value as $T\rightarrow 0$ in all of the cases. We observe that such value is model-dependent, but seems to be strongly constrained by the number of broken symmetries and the interplay between the anomalies.\\

Section \ref{sec:Brokaxial} is devoted to the study of the chiral transport of a broken anomalous $U(1)$ symmetry. At $\xi =0$, we found the result previously pointed out, namely, the value of the conductivity is $1/3$ of that in the unbroken phase, i.e,
\begin{align}
\sigma_{55}(T\rightarrow 0) \approx \frac{8 \kappa}{3}  \mu_5 =\frac{\sigma_{55}(T_c)}{3}\,.
\end{align}
This fact turns out not to be affected when a supervelocity parallel to the momentum is considered. Moreover, as soon as supervelocity is considered, we have two new anomalous effects present: The Chiral Electric Effect and the Chiral Charge Generation Effect. We proposed suitable Kubo formulae for both the CEE and CCGE and computed their value, finding that both become independent from the chemical potential at sufficiently low temperatures. Moreover, their dependence with the superfluid velocity is linear, i.e.,
\begin{align}
\sigma_{CEE}(T\rightarrow 0) \approx \sigma_{CCGE}(T\rightarrow 0) \approx \frac{8 \kappa}{3}  \xi_x\,,
\end{align}

Section \ref{sec:axialvec} deals with two $U(1)$ global symmetries, giving rise to a more rich set of anomalous conductivities with different behaviors once one of the $U(1)$ symmetries gets spontaneously broken. The transport coefficient $\sigma_{55}$ remains the same as in the unbroken phase, but the CMC now acquires a dependence of the vector chemical potential that makes it vanish as we lower the temperature. This result suggests that the charged particles stored in the condensate (forming "cooper pairs") do not contribute to the CMC, which hence vanishes at sufficiently low temperatures. The decrease of the CMC seems to be following a law of the form $\sigma_{CME}/T\sim g/\bar \mu^2$, with $g \approx 2.15$. The scaling of $\sigma_{CME}$ with the axial chemical potential is the usual one, namely $\sigma_{CME} \sim \mu_A$. Finally, the CSE decreases up to $1/2$ the value that it presents in the unbroken phase, yielding
\begin{align}
\sigma_{CSE}(T\rightarrow 0) \approx \frac{6 \kappa}{2}  \mu =\frac{\sigma_{CSE}(T_c)}{2}\,.
\end{align}

These results do not get altered when inducing supervelocity. Furthermore, we observe $\sigma^{V}_{CEE}=\sigma^{A}_{CEE}$, both presenting a qualitative behavior that is similar to the one of Section \ref{sec:Brokaxial}; however the scaling with supervelocity is now
\begin{align}
\sigma^{(V,A)}_{CEE}(T\rightarrow 0)\approx 3 \kappa  \xi_x
\end{align}

Finally, $\sigma^{V}_{CCGE}\ne\sigma^{A}_{CCGE}$ close to the phase transition. At low temperatures both tend to the same value and the same dependence on supervelocity, namely 
\begin{align}
\sigma^{(V,A)}_{CCGE}(T\rightarrow 0) \approx 3 \kappa   \xi_z
\end{align}
\subsection{On the Low temperature behaviour of the Chiral Conductivities}
\label{subsec:Lowt}
\begin{figure}[h] 
\centering
\includegraphics[width=220pt]{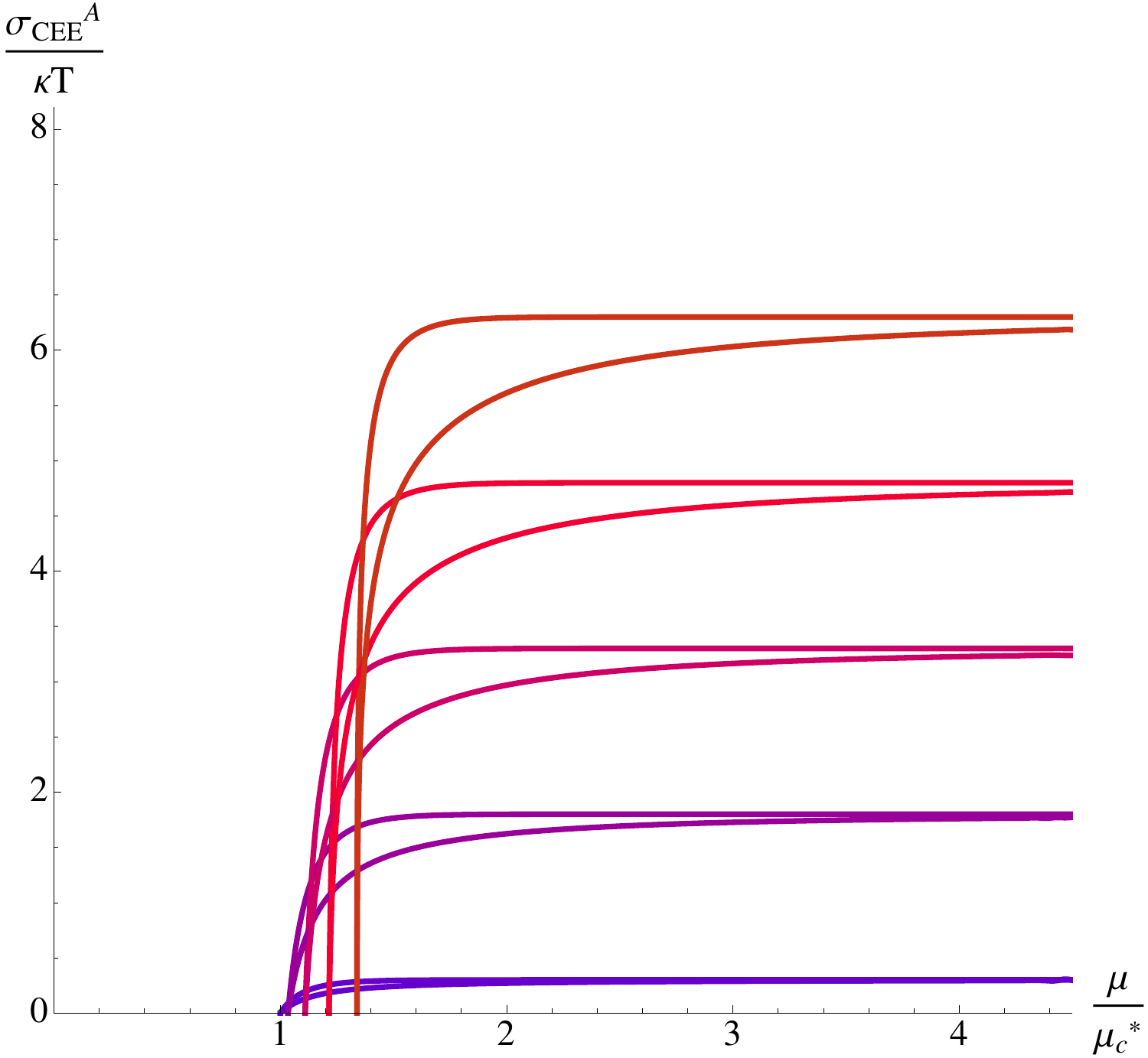}\hspace{2mm}
\includegraphics[width=220pt]{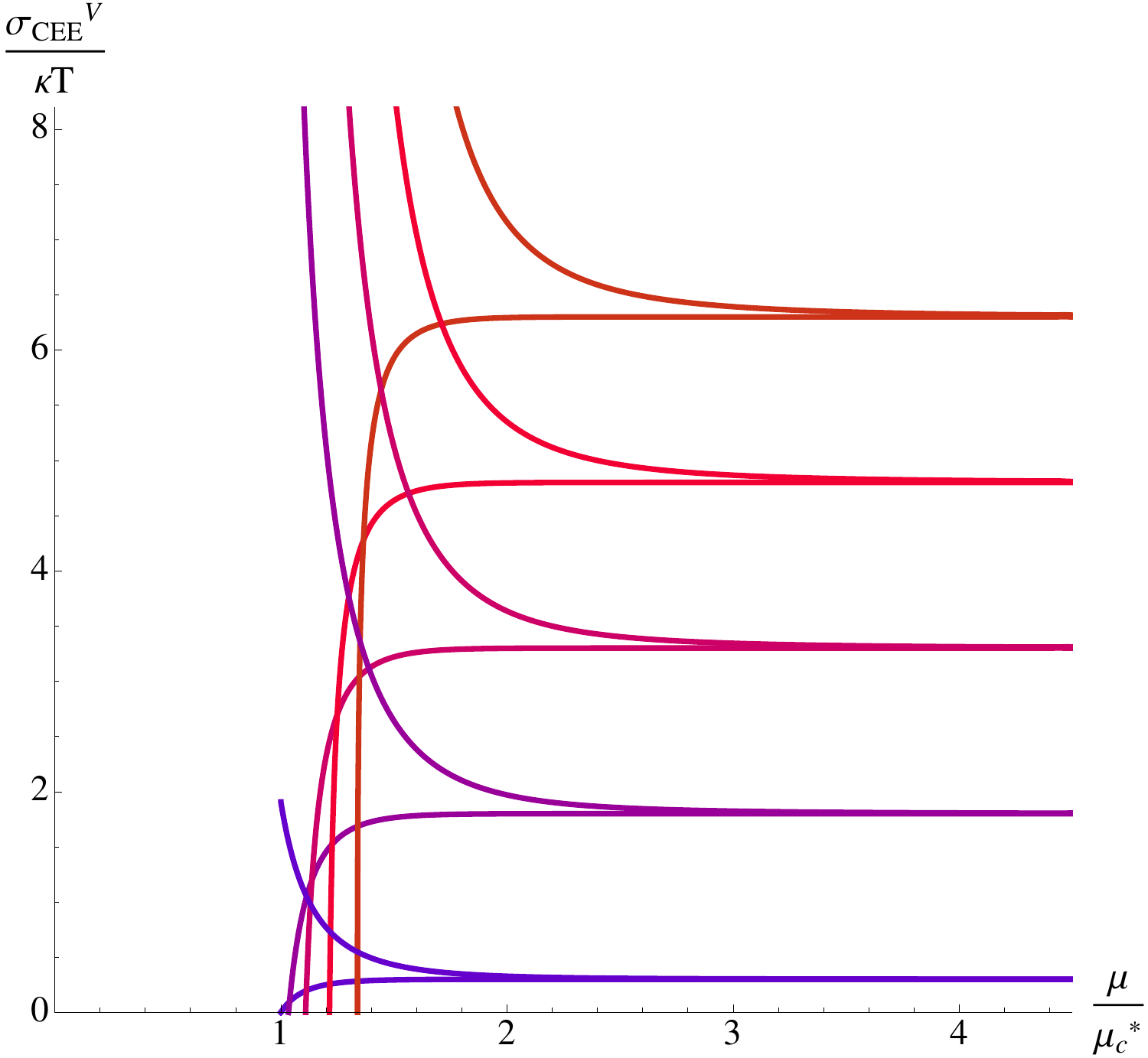}
\caption{\label{fig:ConvCEE} $\sigma^{(A)}_{CEE} /T$ (Left) and $\sigma^{(V)}_{CEE} /T$ (Right) versus vector chemical potential at  $\bar \mu_5= 1$ and $\xi_x/T= 0.1-2.1$ (bottom to top) computed in the two different kinematic limits allowed by gauge invariance. For large enough values of the chemical potential the lines overlap. Notice that one of the limits corresponds to the CCGC.}
\end{figure}

A simple argument by which the CCGE is expected to arise in superfluids is the following. Imagine that we have free Chiral fermions coupled to an electromagnetic field $A_{\mu}$. 
\begin{align}
\mathcal{L} = \bar \psi ( V_{\mu} - A_{\mu})\gamma^{\mu}\psi
\end{align}
We also couple them to an external field $V_{\mu}$ associated to a $U(1)$ symmetry that gets spontaneously broken. The axial current $j^{\mu}_{\text{axial}} = \bar \psi \gamma^{\mu}\gamma^5 \psi$ is anomalous. Hence, in general
\begin{align}
\partial_{\mu} j^{\mu}_{\text{axial}} = a\ F \wedge F + b \ G \wedge F + c\ G \wedge G
\end{align}

where $a, b$ and $c$ are coefficients; $F$ and $G$ are the stress-tensors associated to $A_{\mu}$ and $V_{\mu}$ respectively. There is no external axial field. Let us concentrate on the term proportional to $b$; due to the Bianchi identities, it can be rewritten as $b\ \partial_{\mu} (\epsilon^{\mu \nu \rho \lambda} V_{\nu} F_{\rho \lambda})$\footnote{ Notice that, since the symmetry is spontaneously broken, in principle we have to substitute $V_{\mu} \rightarrow V_{\mu} - \partial_{\mu} \phi$, where $\phi$ is the Goldstone mode. However, for simplicity we stick to a gauge in which $\phi=0$. This will not influence our conclusions.}. At this point, we substitute the actual value of $V_{\mu}$, which, assuming that $\mu=0$, corresponds to $V_{\mu} = (0, \xi_x, 0, 0)$\footnote{One can consider $\xi \rightarrow \xi e^{-i\omega t}$ instead, to bring down the frequency in \ref{currentconc} consistently. At the end of the calculation all the $\omega$ factors will cancel.}. Assuming that $j^{\mu}_{\text{axial}}$ does not depend on the position, we find, in momentum space
\begin{align}
\label{currentconc}\omega j^{\mu}_{\text{axial}} = \omega b \epsilon^{\mu x \rho \lambda} \xi_x F_{\rho \lambda} + ...
\end{align}
leading to both the Chiral Electric Effect and the Chiral Charge Generation Effect, i.e,
\begin{align}
j^{y}_{\text{axial}} \sim b \epsilon^{y x t z} \xi_x E_z\\
\label{CCGEconc} j^{t}_{\text{axial}} \sim b \epsilon^{t x y z} \xi_x B_x
\end{align}
The above argument "with the hands" leads us to some notion of covariantization of those effects\footnote{A cautionary remark is in order here. It is not clear to us whether an argument such as the one presented here gives the complete answer, i.e. whether one can associate the parameter $b$ in (\ref{CCGEconc}) to the actual $\sigma_{CCGE}$. Most likely one cannot. The reason for our concerns is that, for instance, the reasoning does not distinguish between covariant/consistent currents and overlooks the subtleties associated to the introduction of chemical potential/supervelocity in the presence of anomalous symmetries. However, we believe that it serves to ilustrate the kind of transport phenomena that we expect, for it works at the formal level.}. This would imply that for the $U(1)^3$ anomaly, the anomalous contribution to the current can be recast in a covariant form
\begin{align}
\label{currentcov} J^{\mu}_{\text{anom}} (T\rightarrow 0) =  \Sigma^{\mathcal{A}}_{SCE}\epsilon^{\mu \nu \rho \lambda} u^S_{\nu} F_{\rho \lambda} + ...
\end{align}
where $u^{S \mu} = -\mu u^{\mu} +\zeta_{\nu} P^{\nu \mu} $ is the (non-normalized) superfluid velocity and the "$...$" indicate possible corrections due to vorticity. This covariant form of the response can be analyzed numerically by establishing the numerical universality (up to the form of the interplay between the anomaly $\mathcal{A}$ and the broken symmetries) of the coefficient $\Sigma^{\mathcal{A}}_{SCE}$ .  Our results suggest that the superfluid component (the only one present at zero temperature) gives a contribution of the form (\ref{currentcov}) with
\begin{align}
\label{SigmaAAA} \Sigma_{SCE}^{AAA} = \frac{C}{3}
\end{align}
being $C$ a number that is fully determined by the anomaly coefficient.\\
For the $U(1)\times U(1)$ model the at zero temperatures there exists a subset of non-vanishing chiral conductivities for which (\ref{currentcov}) applies, with 
\begin{align}
\label{SigmaAVV} \Sigma_{SCE}^{AVV} = \frac{C}{2}\,.
\end{align}
Equations (\ref{SigmaAAA}) and (\ref{SigmaAVV}) are very suggestive. The nature of the number in the denominator of  $\Sigma^{\mathcal{A}}_{SCE}$ appears to be determined by the spontaneously broken symmetries that are contained in the anomaly responsible for the chiral conductivity under consideration. \\
Furthermore, one could ask whether the conclusions presented here are universal, i.e. valid for all holographic s-wave superfluids or even beyond the holographic approach. If (\ref{SigmaAAA}) and (\ref{SigmaAVV}) held in general, it would imply that at zero temperature the anomalous conductivities have a robust value, entirely determined by anomaly coefficients plus the interplay between the broken symmetries and the anomalies.\\
We would also like to emphasize that formula (\ref{CCGEKubo}) allows us to extract the coefficient termed $g_1(T, \mu/T, \xi^2 /T^2)$ in \cite{Bhattacharyya:2012xi}. At low temperatures, our numerical results for the CCGC and $\sigma_{55}$ for the $U(1)^3$ anomaly are perfectly compatible with
\begin{align}
\label{g1} g_1 (\bar \mu >>1) = - \frac{C}{3}\frac{\mu}{T}
\end{align}
In the case of the AVV anomaly, the compatibility seems to be not that straightforward. \\
In the notation of \cite{Chapman:2013qpa}, $\sigma_{55}\sim (2 T g_1 +\mu C )$. The coefficient $g_1$ is continuous at the phase transition but its derivative is not (see Figure \ref{fig:Sigma55vsV}) and hence $\sigma_{CCGE}\sim g_{1,\nu}$ is not continuous at $\bar \mu_c$. This fact explains why we do not observe that the CCGC vanishes at the phase transition. \\
Finally, let us mention that the electric field $E_x = \partial_{[t} A_{x]}$ is a gauge invariant source in our setup, so assuming that $j^y\sim \sigma_{CEE} E_x$ \emph{only}, we would have expected 
\begin{align}
\label{gaugeinvrel} \frac{i}{\xi_z}\lim_{\omega\rightarrow 0} \partial_{\omega} \mathcal{G}^{yx}_{ra}(\omega, k)|_{k_y=k_x=0} =  \frac{i}{\xi_z}\lim_{k_x\rightarrow 0}  \partial_{k_x}\mathcal{G}^{yt}_{ra}(\omega, k)|_{k_y=k_x=0}
\end{align}
to hold by gauge invariance. Here $\mathcal{G}^{\mu \nu}_{ra}$ are retarded correlators and the subindex "ra" represents the correct combination of time and anti-time ordered sources with respect to which we vary the generating functional. \\
Notice that the right hand side of equation (\ref{gaugeinvrel}) is also the Kubo formula for $\sigma_{CCGE}$, and therefore $\sigma_{CCGE}=\sigma_{CEE}$ should be enforced by gauge invariance of the external sources. This is not what we observe, compare Figures \ref{fig:CECaxialonlyvsmuandxi} and \ref{fig:CGGEvsmu5}. The reason is that the constitutive relation of the current receives contributions from terms other than the one associated to the CEE and therefore the limits taken in (\ref{gaugeinvrel}) capture the influence of gauge-invariant sources that are not the electric field. Remarkably, the effect of those other sources seems to vanish at low temperatures, as shown in Figure \ref{fig:ConvCEE}, for, at $T\rightarrow0$, we recover (\ref{gaugeinvrel}). This supports the validity of the relation (\ref{currentcov}). \\

For future analysis, a possible direction concerns he computation of the Chiral Vortical Conductivity. This amounts to studying the system with backreaction. However, the complicated form of the holographic gauge-gravitational anomaly introduces important difficulties. Moreover, it would also be interesting to analyze the case in which the pattern of broken symmetries is $U(2)\rightarrow U(1)$, for in that case it is known that the spectrum of low-energy excitations is qualitatively different and this could affect the anomalous conductivities.\\

\newpage

\appendix

\section{Computing the Conductivities}

To compute the conductivities we have followed the method developed in \cite{Kaminski:2009dh}.\\

We rearrange the perturbations in a vector $\Phi (r, x^{\mu})$ and work with the Fourier transformed quantity
\begin{align}
\Phi (r, x^{\mu}) = \int \frac{d^d k}{(2 \pi)^d} \Phi_k^I (r) e^{-i\omega t + i \vec k \vec x}
\end{align}

with $\Phi_k (r)$ being
\begin{align} 
\Phi_k^\top (u) = \left(A_t(r), A_x (r),  A_z (r), ... \right) 
\end{align} 

(the specific structure depends on the case at hand, the number of coupled fields, etc.). The general form of the boundary action is \cite{Kaminski:2009dh}
\begin{align} \label{deltact} \delta S^{(2)}=\int \frac{d^dk}{(2\pi)^d}\left[\Phi^I_{-k}\mathcal{A}_{IJ}\Phi'^J_{k}+\Phi^I_{-k}\mathcal{B}_{IJ}\Phi^J_{k}  \right] 
\end{align}
where the prime stands for $d/dr$. To calculate the retarded correlators we solve the equations for the
perturbations with infalling boundary conditions, on the one hand, and boundary conditions $\Phi^I_k (r \rightarrow \infty) = \phi^I_k$ on the other. This procedure should give us the desired Green's functions, after taking the variation of (\ref{deltact}) with respect to the fields at large values of $r$. Moreover, if 
\begin{align}
\Phi^I_{k}(r) = F^I_J(k,r) \phi^J_k
\end{align} 

then $ F^I_J(k,r \rightarrow \infty)=1$ is the bulk-to-boundary propagator. The retarded two-point functions, from which we are able to read directly the transport coefficients, are then computed as
 \begin{align} 
 \label{defcorrel} G^{\mathcal{R}}_{IJ} (k, r\rightarrow \infty) = -2 \lim_{r \rightarrow
\infty } \left(\mathcal{A}_{IM}\left(F^M_J(k,r)\right)'+ \mathcal{B}_{IJ} \right)
\end{align}

The $\mathcal{A}_{IJ}$ and $\mathcal{B}_{IJ}$ matrices depend only on the background and also upon the model under consideration. We provide their values below

\subsection{$U(1)$ model: $\mathcal{A}_{IJ}$ and $\mathcal{B}_{IJ}$ matrices} 

The matrices turn out to be independent of the supervelocity and its direction, once we neglect the contribution of the Chern-Simons term to define the covariant currents. We get
\begin{align} 
\nonumber \mathcal{A} =-\frac{1}{2} r f(r) \text{Diag}(1, 1)\\
\nonumber \mathcal{B}=0\\
\mathcal{B}_{\text{CT}} =\frac{\ln r}{4}\left(\frac{k^2 \sqrt{f(r)}}{r} -\frac{\omega^2 r}{\sqrt{f(r)}}\right) \text{Diag} (1,1)
\end{align}
 
Notice that the counterterms do not contribute to the anomalous transport coefficients, for $\mathcal{B}_{\text{CT}}$ only has diagonal entries, which furthermore are of second order in $\omega$ and $k$.
\subsection{$U(1)\times U(1)$ model: $\mathcal{A}_{IJ}$ and $\mathcal{B}_{IJ}$ matrices} 

In this case we get the same results as before, independently for the axial and vector fields, namely
\begin{align} 
\nonumber \mathcal{A_{\text{axial}}} =-\frac{1}{2} r f(r) \text{Diag}(1, 1)\\
\nonumber \mathcal{B_{\text{axial}}}=0\\
\mathcal{B}^{\text{axial}}_{\text{CT}} =\frac{\ln r}{4}\left(\frac{k^2 \sqrt{f(r)}}{r} -\frac{\omega^2 r}{\sqrt{f(r)}}\right) \text{Diag} (1,1)
\end{align}

and
\begin{align} 
\nonumber \mathcal{A_{\text{vector}}} =-\frac{1}{2} r f(r) \text{Diag}(1, 1)\\
\nonumber \mathcal{B_{\text{vector}}}=0\\
\mathcal{B}^{\text{vector}}_{\text{CT}} =\frac{\ln r}{4}\left(\frac{k^2 \sqrt{f(r)}}{r} -\frac{\omega^2 r}{\sqrt{f(r)}}\right) \text{Diag} (1,1)
\end{align}

\section{Equations of Motion}
\subsection{Momentum transverse to the supervelocity for the $U(1)$ model}\label{primerastransverse}
\begin{align}
 0=f\rho'' +\left(f' +\frac{3f}{r}\right) \rho' +\left(\frac{\omega^2}{f}+\frac{\phi^2}{f}- \frac{ V^2}{r^2}- \frac{k^2}{r^2}  -m^2\right)\rho + \frac{2 i\omega \phi}{f}\delta +2 a_t\Psi\frac{\phi}{f} - 2\frac{ a_z}{r^2}\Psi V \\
 0=f\delta'' +\left(f' +\frac{3f}{r}\right) \delta' +\left(\frac{\omega^2}{f}+\frac{\phi^2}{f}- \frac{ V^2}{r^2} - \frac{k^2}{r^2}  -m^2\right)\delta - \frac{2 i\omega \phi}{f}\rho - i \Psi\omega\frac{a_t}{f} - k \frac{i}{r^2} \Psi a_x\\
0=f a''_t + \frac{3f}{r} a'_t-\left(\frac{k^2}{r^2}+2\Psi^2\right) a_t - \frac{\omega k}{r^2} a_x-4\Psi \phi \rho - 2i\omega\Psi \delta- 16 i k \kappa \frac{f}{r^3} V' a_y\\
0=\label{QNM03} f a''_x +\left(f' + \frac{f}{r}\right) a'_x +\left(\frac{\omega^2 }{f} -2 \Psi^2\right)a_x +\frac{\omega k}{f}a_t+2ik \Psi \delta +\frac{16 i  \kappa}{r} \omega  V' a_y\\
 f a''_y + \left(f' + \frac{f}{r}\right) a_y' + \left(\frac{\omega^2}{f} - \frac{k^2}{r^2} - 2\psi^2\right)a_y + 16 ik \frac{\kappa}{r}\phi' a_z - \frac{16 i  \kappa}{r}V'\left(\omega a_x+ k a_t\right)=0 \\ 
f a''_z + \left(f' + \frac{f}{r}\right) a_z' + \left(\frac{\omega^2}{f} - \frac{k^2}{r^2} - 2\psi^2\right)a_z - 16 ik \frac{\kappa}{r}\phi' a_y -4 V \Psi \rho=0
\end{align}

and the constraint
\begin{align}
\label{constaxial} 0=\frac{i \omega}{f}a'_t +\frac{ik}{r^2}a'_x + 2 \Psi' \delta - 2\Psi \delta'
\end{align}

Where $a_i$ are the perturbations of the axial gauge field. $\rho$ and $\delta$ are the real and imaginary parts of the perturbations of the scalar field, respectively. Momentum points in the $x$-direction, transverse to the superfluid velocity that points in the $z$-direction.
We observe that now the equations become more complicated, with the perturbations of the scalar coupled to all the fields, including the transverse sector. This can imply that the Quasinormal Modes now get affected by the anomaly. 
\subsection{Momentum parallel to the supervelocity for the $U(1)$ model} \label{eqsparalel}
The equations for the relevant sector with momentum aligned to the supervelocity read
\begin{align}
\label{eqnpert1brovec} \nonumber v''_y + \left(\frac{f'}{f} + \frac{1}{r}\right) v'_y + \frac{1}{f}\left(\frac{\omega^2}{f} - \frac{k^2 L^2}{r^2} - 2\psi^2\right)v_y + 12 ik \frac{\kappa L }{rf}\phi' a_z+12 ik \frac{\kappa L }{rf}K' v_z\\+12 i\omega  \frac{\kappa L }{rf}V' a_z =0 \\ 
\label{eqnpert2brovec} \nonumber v''_z + \left(\frac{f'}{f} + \frac{1}{r}\right) v'_z + \frac{1}{f}\left(\frac{\omega^2}{f} - \frac{k^2 L^2}{r^2} - 2\psi^2\right)v_z - 12 ik \frac{\kappa L }{rf}\phi' a_y- 12 ik \frac{\kappa L }{rf}K' v_y\\-12 i\omega  \frac{\kappa L }{rf}V' a_y =0 \\ 
\label{eqnpert1broax} a''_y + \left(\frac{f'}{f} + \frac{1}{r}\right) a_y' + \frac{1}{f}\left(\frac{\omega^2}{f} - \frac{k^2 L^2}{r^2} \right)a_y + 12 ik \frac{\kappa L }{rf}\phi' v_z+ 12 ik \frac{\kappa L }{rf}K' a_z+12 i\omega  \frac{\kappa L }{rf}V' v_z =0 \\ 
\label{eqnpert2broax} a''_z + \left(\frac{f'}{f} + \frac{1}{r}\right) a_z' + \frac{1}{f}\left(\frac{\omega^2}{f} - \frac{k^2 L^2}{r^2} \right)a_z - 12 ik \frac{\kappa L }{rf}\phi' v_y- 12 ik \frac{\kappa L }{rf}K' a_y-12 i\omega  \frac{\kappa L }{rf}V' v_y =0
\end{align}
where $v_{\{y,z\}}$ and $a_{\{y,z\}}$ are the vector and axial perturbations respectively. Momentum points in the $x$-direction, parallel to the supervelocity. Note that only the vector component couples to the condensate, as could have been anticipated. This equations decouple from the equations for the rest of perturbations.
\subsection{Momentum transverse to the supervelocity for the $U(1)\times U(1)$ model}\label{ultimapendix}
The equations read 
\begin{align}
\label{eqnpertrhotransvel}f \rho'' + \left(f' + \frac{3f }{r}\right) \rho+\left( \frac{\omega^2 +\phi^2}{f}- \frac{k^2+V^2}{r^2} -m^2 \right) \rho- \frac{2}{r^2} \psi V v_z +\frac{2\phi}{f}\left(\psi v_t + i\omega \delta \right)=0\\
\label{eqnpertdeltatransvel} f \delta'' + \left(f' + \frac{3f }{r}\right) \delta+\left( \frac{\omega^2 +\phi^2}{f}- \frac{k^2+V^2}{r^2} -m^2 \right) \delta- \frac{i}{r^2} \psi k v_x -\frac{i\omega}{f}\left(\psi v_t + 2 \phi \rho \right)=0\\
\label{eqnpertvttransvel} f v''_t +  \frac{3f}{r} v'_t - \left(\frac{k^2}{r^2}+ 2 \psi^2\right) v_t - \frac{\omega k}{r^2}v_x  - 2i \omega \psi  \delta - 4 \phi \psi \rho - 12 ik \frac{\kappa f}{r^3}V' a_y =0 \\ 
\label{eqnpertvxtransvel} f v''_x +  \left( f' + \frac{f}{r} \right)v'_x+\left(\frac{\omega^2}{f}- 2 \psi^2\right) v_x + \frac{\omega k}{f}v_t  + 2i k \psi  \delta + 12 i\omega \frac{\kappa}{r}V' a_y =0 \\ 
\label{eqnpertvytransvel}\nonumber f v''_y + \left(f' + \frac{f}{r}\right) v'_y + \left(\frac{\omega^2}{f} - \frac{k^2 }{r^2} - 2\psi^2\right)v_y + 12 ik \frac{\kappa  }{r}\phi' a_z+ 12 ik \frac{\kappa}{r}K' v_z-  \\ 12 i\omega \frac{\kappa}{r}V' a_x-24 ik \frac{\kappa}{r}V' a_t =0 \\ 
\label{eqnpertvztransvel} f v''_z + \left(f' + \frac{f}{r}\right) v'_z + \left(\frac{\omega^2}{f} - \frac{k^2 }{r^2} - 2\psi^2\right)v_z - 4 V \psi \rho - 12 ik \frac{\kappa  }{r}\phi' a_y- 12 ik \frac{\kappa}{r}K' v_y =0 \\ 
\label{eqnpertattransvel} f a''_t +  \frac{3f}{r} a'_t - \frac{k^2}{r^2}a_t - \frac{\omega k}{r^2}a_x  - 12 ik \frac{\kappa f}{r^3}V' v_y =0 \\ 
\label{eqnpertaxtransvel} f a''_x +  \left( f' + \frac{f}{r} \right)a'_x+\frac{\omega^2}{f}a_x+ \frac{\omega k}{f}a_t  + 12 i\omega \frac{\kappa}{r}V' v_y =0 \\ 
\label{eqnpertaytransvel}\nonumber f a''_y + \left(f' + \frac{f}{r}\right) a'_y + \left(\frac{\omega^2}{f} - \frac{k^2 }{r^2}\right)a_y + 12 ik \frac{\kappa  }{r}\phi' v_z+ 12 ik \frac{\kappa}{r}K' a_z-  \\ 12 i\omega \frac{\kappa}{r}V' v_x-12 ik \frac{\kappa}{r}V' v_t =0 \\ 
\label{eqnpertaztransvel} f a''_z + \left(f' + \frac{f}{r}\right) a'_z + \left(\frac{\omega^2}{f} - \frac{k^2 }{r^2} \right)a_z  - 12 ik \frac{\kappa  }{r}\phi' v_y=0  
\end{align}
And the constraints
\begin{align}
\frac{i \omega}{f}a'_t +\frac{ik}{r^2}a'_x=0 \\
\frac{i \omega}{f}v'_t +\frac{ik}{r^2}v'_x + 2 \psi' \delta - 2\psi \delta'=0
\end{align}
Where $a_i$ and $v_i$ are the perturbations of the axial and vector gauge fields respectively. $\rho$ and $\delta$ are the real and imaginary parts of the perturbations of the scalar field, respectively. Momentum points in the $x$-direction, whereas the superfluid velocity points in the $z$-direction.

\section*{Acknowledgments}
We would like to thank K.Landsteiner especially for illuminating discussions and constant support. We also want to thank C.Hoyos and I.Amado for useful comments on the draft.  A. J. and L. M. are supported by Plan Nacional de Altas Energ\'\i as FPA
2009-07890, Consolider Ingenio 2010 CPAN CSD200-00042 and
HEP-HACOS S2009/ESP-2473. L.M. has been supported by FPI-fellowship
BES-2010-041571. A. J. has been supported by FPU fellowship AP2010-5686. L.M. thanks Susana Hernández for tremendously joyful discussions.

\newpage

\addcontentsline{toc}{section}{References}

\nocite{*}

\bibliographystyle{jhepcap}
\bibliography{Anomsupv2}

\end{document}